\documentclass[aps,pre,groupedaddress,twocolumn,floatfix]{revtex4-1}
\usepackage{epsfig}
\usepackage{natbib}

\begin{document}
\newcommand{\h}{\phi}
\newcommand{\mon}{\begin{displaymath}}
\newcommand{\moff}{\end{displaymath}}
\newcommand{\sumi}[1]{\sum_{{#1}=-\infty}^{\infty}}
\renewcommand{\b}[1]{\mbox{\boldmath ${#1}$}}
\newcommand{\sumy}{\sum_{\b{y}}}
\newcommand{\sumz}{\sum_{\b{z}}}
\newcommand{\pd}[2]{\frac{\partial {#1}}{\partial {#2}}}
\newcommand{\od}[2]{\frac{d {#1}}{d {#2}}}
\newcommand{\inti}{\int_{-\infty}^{\infty}}
\newcommand{\eon}{\begin{equation}}
\newcommand{\eoff}{\end{equation}}
\newcommand{\eaon}{\begin{eqnarray}}
\newcommand{\eaoff}{\end{eqnarray}}
\newcommand{\e}[1]{\times 10^{#1}}
\newcommand{\chem}[2]{{}^{#2} \mathrm{#1}}
\newcommand{\s}{s}
\newcommand{\zetaexp}{\left( \zeta e^{q \s t} \right)}
\newcommand{\taunuc}{\tau_{nuc}}
\newcommand{\eq}[1]{Eq. (\ref{#1})}
\newcommand{\ev}[1]{\left\langle #1 \right\rangle}
\newcommand{\mat}[1]{\bf{\mathcal{#1}}}
\newcommand{\fig}[1]{Fig. \ref{#1}}
\newcommand{\ub}{U_b}
\newcommand{\dcoal}[2]{D_{#1#2}}
\newcommand{\ph}{\dcoal{H}{1}}
\newcommand{\Prob}{\mathrm{Pr}}
\renewcommand{\H}{H}
\newcommand{\Co}{{\cal{C}}}
\newcommand{\Gam}{\Gamma}
\newcommand{\la}{\langle}
\newcommand{\ra}{\rangle}
\newcommand{\Oh}{{\cal O}}
\newcommand{\pib}[2]{\pi^b_{#1, #2}}
\newcommand{\un}{U_n}

\newcommand{\tsw}{\tau_{nm}}
\newcommand{\tmrca}{J}

\renewcommand{\baselinestretch}{1.0}

\title{Genetic Diversity and the Structure of Genealogies in Rapidly Adapting Populations}
\author{Michael M. Desai$^{1*}$}
\author{Aleksandra M. Walczak$^{2*}$}
\author{Daniel S. Fisher$^{3}$}
\affiliation{\mbox{${}^1$Department of Organismic and Evolutionary Biology, Department of Physics, and} \mbox{FAS Center for Systems Biology, Harvard University} \\ \mbox{${}^{2}$CNRS-Laboratoire de Physique Th\'eorique de l'\'Ecole Normale Sup\'erieure,} \\ \mbox{${}^3$Department of Applied Physics and Department of Bioengineering, Stanford University} \\ \mbox{${}^*$These authors contributed equally to this work} }

\begin{abstract}

Positive selection distorts the structure of genealogies and hence alters patterns of genetic variation within a population.  Most analyses of these distortions focus on the signatures of hitchhiking due to hard or soft selective sweeps at a single genetic locus.  However, in linked regions of rapidly adapting genomes, multiple beneficial mutations at different loci can segregate simultaneously within the population, an effect known as clonal interference.  This leads to a subtle interplay between hitchhiking and interference effects, which leads to a unique signature of rapid adaptation on genetic variation both at the selected sites and at linked neutral loci.  Here, we introduce an effective coalescent theory (a ``fitness-class coalescent'') that describes how positive selection at many perfectly linked sites alters the structure of genealogies.  We use this theory to calculate several simple statistics describing genetic variation within a rapidly adapting population, and to implement efficient backwards-time coalescent simulations which can be used to predict how clonal interference alters the expected patterns of molecular evolution.

\end{abstract}

\date{\today}

\maketitle

\section{Introduction}

Beneficial mutations drive long-term evolutionary adaptation, and despite their rarity they can dramatically alter the patterns of genetic diversity at linked sites.  Extensive work has been devoted to characterizing these signatures in patterns of molecular evolution, and using them to infer which mutations have driven past adaptation.

When beneficial mutations are rare and selection is strong, adaptation progresses via a series of selective sweeps.  A single new beneficial mutation occurs in a single genetic background, and increases rapidly in frequency towards fixation.  This is known as a ``hard'' selective sweep, and it purges genetic variation at linked sites and shortens coalescence times near the selected locus \citep{maynardsmithhaigh74}.  Most statistical methods used to detect signals of adaptation in genomic scans are based on looking for signatures of these hard sweeps \citep{sabeti06, akey09, nielsenclark07, novembredirienzo09, pritchardcoop10}.

\begin{figure*}
\includegraphics[width=6.5in]{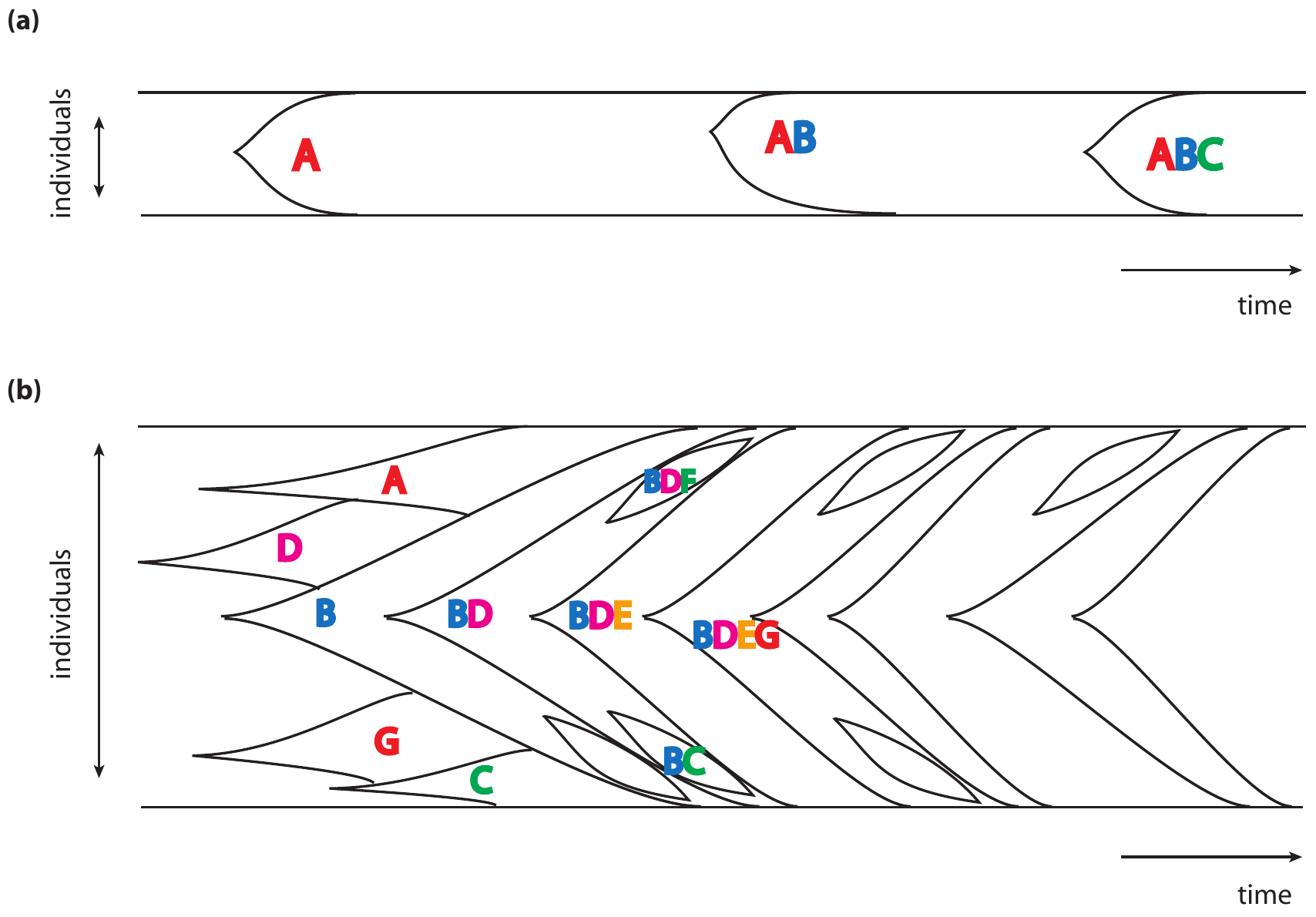}
\caption{Schematic of the evolutionary dynamics of adaptation.  (\textbf{a}) A small population adapts via a sequence of selective sweeps.  (\textbf{b}) In a large rapidly adapting population, multiple beneficial mutations segregate concurrently.  Some of these mutant lineage interfere with each others' fixation, while others hitchhike together.  \label{fig1} }
\end{figure*}

Hard selective sweeps are the primary mode of adaptation in small to moderate sized populations in which beneficial mutations are sufficiently rare.  However, in larger populations where beneficial mutations occur more frequently, many different mutant lineages can segregate simultaneously in the population.  If the loci involved are sufficiently distant that recombination occurs frequently enough between them, their fates are independent and adaptation will proceed via independent hard sweeps at each locus.  However, in largely asexual organisms such as microbes and viruses, and on shorter distance scales within sexual genomes, selective sweeps at linked loci can overlap and interfere with one another.  This is referred to as clonal interference, or Hill-Robertson interference in sexual organisms \citep{gerrishlenski98, hillrobertson66}.  These interference effects can dramatically change both the evolutionary dynamics of adaptation and the signatures of positive selection in patterns of molecular evolution.  We illustrate them schematically in \fig{fig1}.

We and others have characterized the evolutionary dynamics by which a population accumulates beneficial mutations in the presence of clonal interference \citep{kesslerlevine98, gerrishlenski98, rouzine03, desaifisher07, hallatschek11, good12}.  Many recent experiments in a variety of different systems have confirmed that these interference effects are important in a wide range of laboratory populations of microbes and viruses \citep{devisser99, miralles99, bollbackhuelsenbeck07, desaifishermurray07, kaosherlock08}.  These theoretical and experimental developments have recently been reviewed by \citet{park10} and \citet{sniegowski10}.

Although this earlier theoretical work has provided a detailed characterization of evolutionary dynamics in the presence of clonal interference, it does not make any predictions about the patterns of genetic variation within an adapting population.  In this paper, we address this question of how clonal interference alters the structure of genealogies, and how this affects patterns of molecular evolution both at the sites underlying adaptation and at linked neutral sites.  This has become particularly relevant in light of recent advances that now make it possible to sequence individuals and pooled population samples from microbial adaptation experiments \citep{barrick09a, barrick09b, gresham08, kaosherlock08}.

We note that much recent work in molecular evolution and statistical genetics has analyzed related scenarios where adaptation involves multiple mutations, motivated by recent theoretical work \cite{ralphcoop10, innankim04, orrbetancourt01} and empirical data from \emph{Drosophila} \cite{sellaandolfatto09} and humans \cite{cooppritchard09, hernandezprzeworski11} that suggests that simple hard sweeps may be rare.  This includes most notably analysis of the effects of ``soft sweeps,'' where recurrent beneficial mutations occur at a single locus, or selection acts on standing variation at this locus \citep{hermissonpennings05, penningshermisson06a, penningshermisson06b}.  Soft sweeps drive multiple genetic backgrounds to moderate frequencies, leaving several deeper coalescence events and hence a weaker signature of reduced variation in the neighborhood of the selected locus than a hard sweep \citep{przeworskiwall05}.

In contrast to the situation we analyze here, both hard and soft sweeps refer to the action of selection at a single locus.  We consider instead a case more analogous to models in quantitative genetics, where selection acts on a large number of loci that all affect fitness.  In other words, our analysis of clonal interference can be thought of as a description of polygenic adaptation, where selection favors the individuals who have beneficial alleles at multiple loci.  Recent work has argued for the potential importance of polygenic adaptation from standing genetic variation \citep{pritcharddirienzo10, pritchardcoop10}, loosely analogous to the case where soft sweeps act at many loci simultaneously \cite{chevinhospital08, hancockdirienzo10}.  Our analysis in this paper, by contrast, describes polygenic adaptation via multiple new mutations of similar effect at many loci, where each locus has a low enough mutation rate that it would undergo a hard sweep in the absence of the other loci.

As with hard and soft sweeps, the signatures of this form of adaptation on nearby genomic regions are determined by how it alters the structure and timing of coalescence events.  In this paper, we therefore focus on computing how clonal interference alters the structure of genealogies.  This involves two basic effects.  On the one hand, mutations at the many loci occur and segregate simultaneously, interfering with each others' fixation.  This preserves some deeper coalescence events, as in a soft sweep.  On the other hand, since the mutations occur at different sites, multiple beneficial mutations can also occur in the same genetic background and hitchhike together.  This tends to shorten coalescence times, making the signature of adaptation somewhat more like a ``hard sweep.''  Together, these effects lead to unique patterns of genetic diversity characteristic of clonal interference.

Our analysis of these effects is based on the fitness-class coalescent we previously used to describe the effects of purifying selection on the structure of genealogies \citep{negselcoalescent}.  This in turn is closely related to the structured coalescent model of \citet{hudsonkaplan94}.  We begin in the next section by describing our model, and summarize our earlier analysis of the rate and dynamics of adaptation in the presence of clonal interference, which describes the distribution of fitnesses within the population \citep{desaifisher07}.  We then show how one can trace the ancestry of individuals as they ``move'' between different fitness classes via mutations (our fitness-class coalescent approach).  We compute the probability that any set of individuals coalesce when they are within the same fitness class.  This leads to a description of the probability of any possible genealogical relationship between a sample of individuals from the population.  Finally, we show how the distortions in genealogical structure caused by clonal interference alter the distributions of simple statistics describing genetic variation at the selected loci as well as linked neutral loci.  We also use our approach to implement coalescent simulations analogous to those previously used to describe the action of purifying selection \citep{gordocharlesworth02, seger10}, based on the structured coalescent method of \citet{hudsonkaplan94}.  These coalescent simulations can be used to analyze in detail how this form of selection alters the structure of genealogies.

Our results provide a theoretical framework for understanding the patterns of genetic diversity within rapidly evolving experimental microbial populations.  Our analysis may also have relevance for understanding how pervasive positive selection alters patterns of molecular evolution more generally, but we emphasize that our work here focuses entirely on asexual populations or on diversity within a short genomic region that remains perfectly linked over the relevant timescales.  In the opposite case of strong recombination, adaptation will progress via independent hard selective sweeps at each selected locus.  Further work is required to understand the effects of intermediate levels of recombination, where the approach recently introduced by \citet{neher10} may provide a useful starting point.

\section{Model and Evolutionary Dynamics}

\subsection{Model}

We consider a finite haploid asexual population of constant size $N$, in which a large number of beneficial mutations are available, each of which increases fitness by the {\it same} amount $s$.  We define $U_b$ as the total mutation rate to these mutations.  We neglect deleterious mutations and beneficial mutations with other selective advantages.  We have previously shown that the dynamics in rapidly adapting populations are dominated by beneficial mutations of a specific fitness effect \citep{desaifisher07,nagle08, good12}, so this model is a useful starting point, but we return to discuss these assumptions further in the Discussion. We also assume that there is no epistasis for fitness, so the fitness of an individual with $k$ beneficial mutations is $w_k = (1+s)^k \approx 1 + s k$.  This is the same model of adaptation we have previously considered \citep{desaifisher07} and is largely equivalent to models used in most related theoretical work on clonal interference \citep{rouzine03, rouzine08, park10}.  We will later also consider linked neutral sites with total mutation rate $\un$, but for now we focus on the structure of genealogies and neglect neutral mutations.

To analyze expected patterns of genetic variation, we must also make specific assumptions about how mutations occur at particular sites. We will consider a perfectly linked genomic region which has a total of $B$ loci at which beneficial mutations can occur.  We assume these mutations occur at rate $\mu$ per locus, for a total beneficial mutation rate $\ub = \mu B$.  We will later take the infinite-sites limit, $B \to \infty$, while keeping the overall beneficial mutation rate $\ub$ constant.  Each mutation is assumed to confer the same fitness advantage $s$, where $s \ll 1$.  We will also assume throughout that selection is strong compared to mutations, $s \gg \ub$, which allows us to use our earlier results in \citet{desaifisher07} as a basis for our analysis.  Analysis of the opposite case where $s < \ub$ remains an important topic for future work, which could be based on alternative models of the dynamics such as the approach of \citet{hallatschek11}.  Although our model is defined for haploids, our analysis also applies to diploid populations provided that there is no dominance (i.e., being homozygous for the beneficial mutation carries twice the fitness benefit as being heterozygous).

This model is the simplest framework that captures the effects of positive selection on a large number of independent loci of similar effect.  However, the dynamics of adaptation in this model can be complex.  Beginning from a population with no mutations at the selected loci, there is first a transient phase while variation at these loci initially increases.  There is then a steady state phase during which the population continuously adapts towards higher fitness.  Finally, adaptation will eventually slow down as the population approaches a well-adapted state.  In this paper, we focus on the second phase of rapid and continuous adaptation, which has been the primary focus of previous work by us and others \citep{desaifisher07, rouzine08, hallatschek11, park10}.  Our goal is to understand how this continuous rapid adaptation alters the structure of genealogies and hence patterns of genetic variation.  We begin in the next subsection by summarizing the relevant aspects of our earlier results for the distribution of fitness within the population.

\subsection{The distribution of fitness within the population}

In our model in which all beneficial mutations confer the same advantage, $s$, the distribution of fitnesses within the population can be characterized by the fraction of the population, $\h_k$, that has $k$ beneficial mutations more or less than the population average.  We refer to this as ``fitness class $k$.''

When $N$ and $\ub$ are small, it is unlikely that a second beneficial mutation will occur while another is segregating.  Hence adaptation proceeds by a succession of selective sweeps.  In this regime, beneficial mutations destined to survive drift arise at rate $N \ub s$ and then fix in $\frac{1}{s} \ln [ N s ]$ generations.  Thus adaptation will occur by successive sweeps provided that \eon N \ub \ll \frac{1}{\ln [N s]}. \eoff  When this condition is met, the population is almost always clonal or nearly clonal except during brief periods while a selective sweep is occurring.  Thus we will have $\h_0 = 1$ and $\h_k = 0$ for $k \neq 0$.

\begin{figure*}
\includegraphics[width=6.5in]{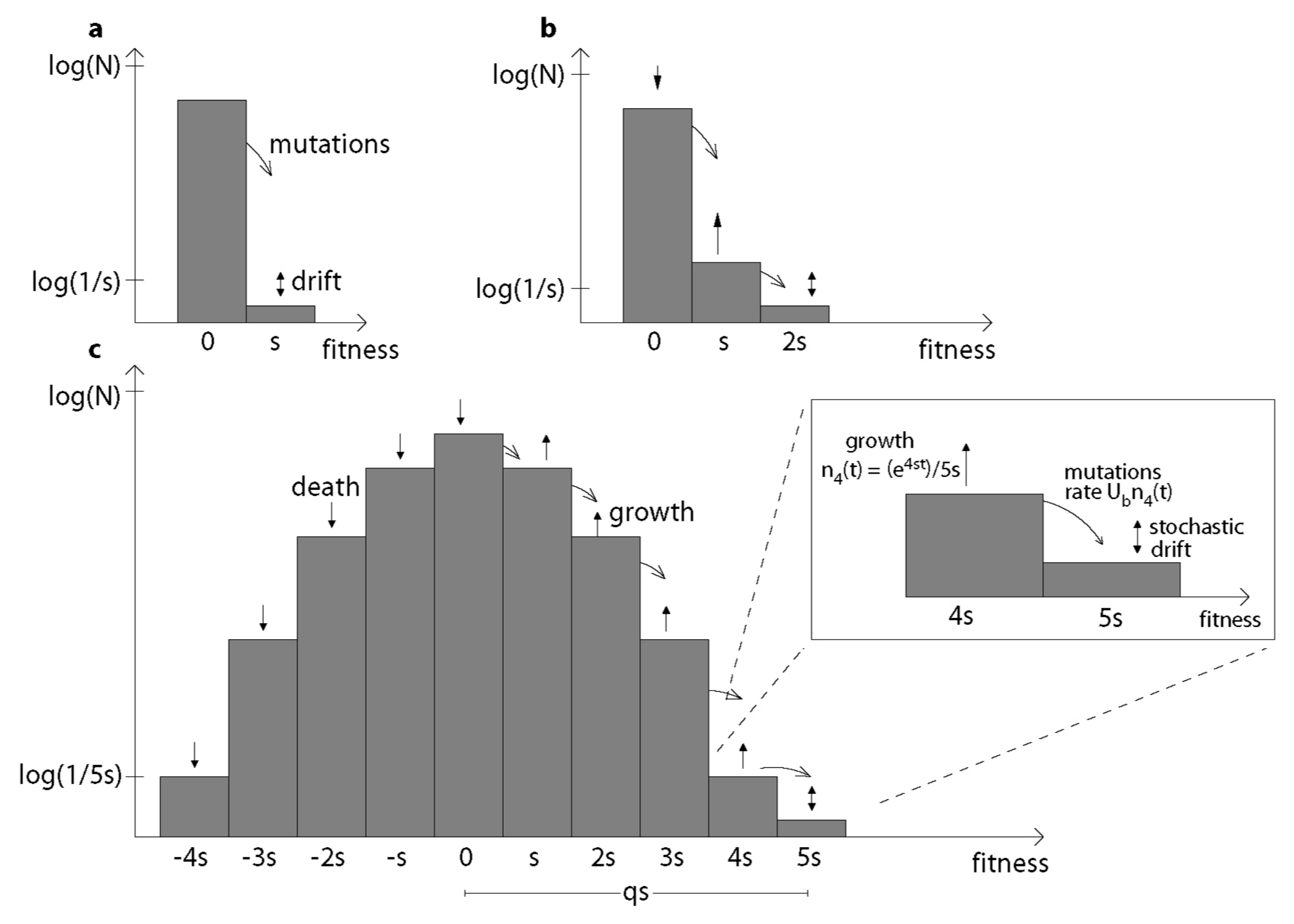}
\caption{Schematic of the evolution of large asexual populations, from \citet{desaifisher07}. The fitness distribution within a population is shown on a logarithmic scale.  (\textbf{a}) The population is initially clonal. Beneficial mutations of effect $s$ create a subpopulation at fitness $s$, which drifts randomly until it reaches a size of order $\frac{1}{s}$, after which it behaves deterministically. (\textbf{b}) This subpopulation generates mutations at fitness $2s$. Meanwhile, the mean fitness of the population increases, so the initial clone begins to decline. (\textbf{c}) A steady state is established. In the time it takes for new mutations to arise, the less fit clones die out and the population moves rightward while maintaining an approximately constant lead from peak to nose, $qs$ (here $q=5$). The inset shows the leading nose of the population. \label{fig2} }
\end{figure*}

In larger populations, however, new mutations continuously arise before the older mutants fix.  Thus the population maintains some variation in fitness even while it adapts.  The distribution of fitnesses within the population is determined by the balance between two effects.  On the one hand, new mutants arise at the high-fitness ``nose'' of this distribution, generating new mutants more fit than any other individuals in the population.  This increases the variation in fitness in the population. (While new mutations occur throughout the fitness distribution, the mutations essential to maintain variation are those that arise at the nose and generate new most-fit individuals.)  On the other hand, selection destroys less-fit variants, increasing the mean fitness and decreasing the variation in fitness within the population.  This is illustrated in \fig{fig2}.

We showed in previous work that this balance between mutation and selection leads to a constant steady state distribution of fitnesses within the population, measured relative to the current (and constantly increasing) mean fitness \citep{desaifisher07}.  In this steady state distribution, the fraction of individuals with $k$ beneficial mutations relative to the current mean in the population is typically \eon \h_k = \h_{-k} = C e^{-\sum_{i=1}^k i s \bar \tau}, \label{hkform} \eoff where $\bar \tau$ is defined below and $C$ is an overall normalization constant that will not matter for our purposes.  Note that the distribution $\h_k$ is approximately Gaussian.

This distribution $\h_k$ is cut off above some finite maximum $k$ which corresponds to the nose of the distribution, the most-fit class of individuals.  We define the \emph{lead} of the fitness distribution, $qs$, as the difference between the mean fitness and the fitness of these most-fit individuals (so $q$ is the maximum value of $k$; the most-fit individuals have $q$ more beneficial mutations than the average individual).  In \citet{desaifisher07}, we showed that \eon q = \frac{2 \ln \left[ N s \right]}{\ln \left[ s/\ub \right]} \approx e^{-(s\bar{\tau}k)^2/2}. \eoff  This is illustrated in \fig{fig2}.

Above we have implicitly defined $\bar \tau$ to be the ``establishment time,'' the average time it takes for new mutations to establish a new class at the nose of the distribution, \eon \bar \tau = \frac{\ln^2 \left[ s / \ub \right]}{2 s \ln \left[ N s \right]}. \eoff  As we will see below, the characteristic time scale for coalescent properties will turn out to be the time for the fitness class at the nose to become the dominant population --- i.e. for the mean fitness to increase by the lead of the fitness distribution.  This takes $q$ establishment times, so that the this ``nose-to-mean'' time is \eon \tsw \approx q \bar{\tau} \approx \frac{\ln(s/U_b)}{s} ,  \eoff which is roughly independent of the population size for sufficiently large $N$.  We note that no single mutant sweeps to fixation in this time: rather, a whole set of mutants comprising a new fitness class at the nose will come to dominate the population a time $\tsw$ later.

\section{The Fitness-Class Coalescent Approach}

We now wish to understand the patterns of genetic variation within a rapidly adapting population in the clonal interference regime.  To do so, we will use a fitness-class coalescent method in which we trace how sampled individuals descended from individuals in less-fit classes, moving between classes by mutation events.  In each fitness class, there will be some probability of coalescence events.  To calculate these coalescence probabilities, we must first understand the clonal structure within each fitness class: this we now consider.

\subsection{Clonal structure}

\begin{figure}
\includegraphics[width=3.4in]{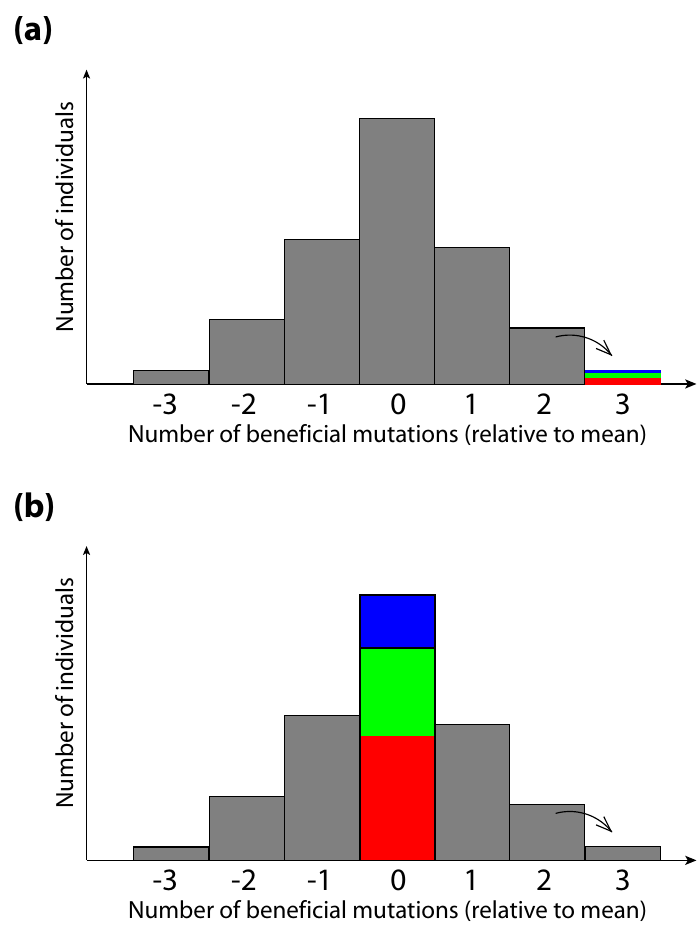}
\caption{Schematic of the establishment and fate of clonal lineages in a given fitness class, shown for a case where $q=3$.  (\textbf{a}) Three new clonal lineages (denoted in different colors) are established at the nose of the fitness distributions by three independent new mutations.  These lineages have relative frequencies determined by the timing of these mutations.  (\textbf{b}) After the population evolves for some time, the class that was at the nose of the distribution in (\textbf{a}) is now at the mean fitness.  The class is still dominated by the three clonal lineages established while the class was at the nose (subsequent mutations represent only a small correction).  These three clonal lineages have the same relative frequencies as when they were established at the nose; these relative frequencies remain ``frozen'' even as the population adapts.  \label{fig3} }
\end{figure}

Each fitness class is first created when a new beneficial mutation occurs in the current most-fit class, creating a new most-fit class at the nose of the fitness distribution (see inset of \fig{fig2}).  This new clonal mutant lineage fluctuates in size due to the effects of genetic drift and selection before it eventually either goes extinct or establishes (i.e. reaches a large enough size that drift becomes negligible).  After establishing, the lineage begins to grow almost deterministically.  Concurrently additional mutations occur at the nose of the distribution, also founding new mutant lineages within this most-fit class.  This process is illustrated in \fig{fig3}a.

We wish to understand the frequency distribution of these new clonal lineages, each founded by a different beneficial mutation.  In our infinite-sites model, each such lineage is genetically unique.  We can gain an intuitive 

understanding of this frequency distribution with a simple heuristic argument.  After it establishes, the size of the current most-fit class, $n_{q-1}(t)$, grows approximately deterministically according to the formula \eon n_{q-1}(t) = \frac{1}{qs} e^{(q-1)st}, \eoff as we described in \citet{desaifisher07}.  New mutations occur in individuals in this class at rate $\ub n_{q-1}(t)$, creating even more-fit individuals.  Each new mutation has a probability $qs$ of escaping genetic drift to form a new established mutant lineage.  Thus the $\ell^{th}$ established mutant lineage at the nose will on average occur at roughly the time $t_\ell$ that satisfies \eon \int_0^{t_\ell} q s \ub n_{q-1}(t) dt = \ell. \eoff  Solving this for $t_\ell$ and then noting that the size, $n_\ell$, of the $\ell^{th}$ established lineage will be proportional to $e^{qs(t-t_\ell)}$, we immediately find \eon \frac{n_\ell}{n_1} \approx \frac{1}{\ell^{1+1/q}} . \eoff  This provides a good estimate of the typical frequency distribution of clonal lineages within this fitness class at the nose, each lineage founded by a single new mutation.

The analysis above describes the clonal structure created as a new fitness class is formed, advancing the nose.  After approximately $\bar \tau$ generations, the mean fitness of the population will have increased by $s$, and the growth rates of all the fitness classes we have described will decrease correspondingly.  Thus we can strictly only use the calculations above up to some finite number of mutations, $\ell_{max}$, after which all growth rates will have decreased due to the advance of the mean fitness of the population. Mutations will continue to occur after this time, but their frequency distribution will be slightly different.  Fortunately, in the strong selection regime we consider ($s \gg \ub$), the total contribution of all mutations after this point to the total size of the class is small compared to the contributions of the mutations that occur while this class is at the nose \citep{desaifisher07, brunetrouzinewilke08}.  We therefore neglect this cutoff to the number of mutations that occur at the nose, as well as the contribution of later mutations. This approximation will break down for very large samples.  However, the errors it introduces can be shown to be relatively small even when considering quantities such as the time to the most recent common ancestor of the whole population.

Another important aspect of the dynamics that simplifies the behavior is that despite the changing growth rate of the fitness class as a whole, the  {\it frequencies} of the established lineages within the class remain fixed.  In other words, the clonal structure within the class remains ``frozen'' after it is initially created, rather than fluctuating with time (see \fig{fig3}b).  As we will see, this and the neglect of late-arising mutations are good approximations in the regimes we consider here.

While our heuristic analysis provides a good picture of the typical frequency distribution of clonal lineages within each fitness class, it misses a crucial effect. Occasionally a new mutation  at the nose will, by chance, occur anomalously early.  This single mutant lineage can then dominate its fitness class.  These events are quite rare, but when they do occur this single lineage can purge a substantial fraction of the total genetic diversity within the population.  As we will see, these events together with less-rare but still early mutations are essential to the understanding the structure of genealogies within the population as they lead to a substantial probability of ``multiple merger'' coalescent events.

To capture these effects, we must carry out a more careful stochastic analysis of the clonal structure within each fitness class.  As before, we focus on the clonal structure created when that class was at the nose of the fitness distribution, since it remains ``frozen'' thereafter.  To do so, we note that the population size at the nose can be written as \eon n = \bar{n}(t)\sum_i \nu_i(t), \eoff where $\bar{n}(t)$ reflects the average growth of all clones due to selection, and $\nu_i(t)$ reflects the stochastic effects of a clone generated from mutations at site $i$ (of $B$ total possible sites).  At late enough times, the distribution of $\nu_i$ becomes time-independent, as shown previously \citep{desaifisher07}.  This time-independent $\nu_i$ summarizes the combined effect of all the stochastic dynamics of mutations at this site that are relevant for the long-term dynamics.  We showed that the generating function of $\nu_i$ is \eon G_i(z) = \ev{e^{-z \nu_i}} = \exp \left[- \frac{1}{B}z^{1-1/q} \right] = e^{-z^\alpha/B} = \left[ 1 - \frac{z^\alpha}{B} \right], \label{two} \eoff where angle brackets denote expectation values, the last equality follows for large $B$, and we have defined \eon \alpha \equiv 1 - \frac{1}{q}.  \eoff  The total size of this fitness class is proportional to \eon \sigma \equiv \sum_{i=1}^B \nu_i.  \eoff

This generating function $G_i(z)$ for the size of the clonal lineage founded at each possible site contains all of the relevant information about the lineage frequency distribution, including the stochastic effects described above.  Below we will use it to calculate coalescence probabilities within our fitness-class coalescent approach, which we now turn to.

\subsection{Tracing Genealogies}

To calculate the structure of genealogies, we take a fitness-class approach analogous to the one we used to analyze the case of purifying selection \citep{negselcoalescent}.  We first consider sampling several individuals from the population.  These individuals come from some set of fitness classes  with probabilities given by the frequencies of those fitness classes, $\h_k$.  We note that in the purifying selection case, fluctuations in the $\h_k$ due to genetic drift were a potential complication in determining these sampling probabilities.  Here, these fluctuations are much less important provided that $\ub/s \ll 1$.  We note however that fluctuations in different $\h_k$ are correlated due to the stochasticity at the nose.  Furthermore, averages of $\h_k$ are far larger than their typical values due to rare fluctuations.  Such fluctuations,  which we discuss in detail elsewhere \citep{dsfjstat}, may lead to some slight corrections to our results. But for most purposes, the typical values of the $\h_k$ are what matters: thus we make the simple approximation that the probability of sampling one individual from class $k_1$ and a second from class $k_2$ is simply $\h_{k_1} \h_{k_2}$, with $\h_k$ as given in \eq{hkform}.  Analogous formulas apply for larger samples.

Each sampled individual comes from a specific fitness class $k$, and belongs to a specific clonal lineage within that class.  This clonal lineage was created when this fitness class was at the nose of the distribution, approximately $(q-k) \bar \tau$ generations ago.  It was created by a single new mutation in an individual from what is now fitness class $k-1$.  That individual in turn belonged to some clonal lineage within class $k-1$, which in turn was created when that class was at the nose by a new mutation in an individual from what is now fitness class $k-2$, and so on.

\begin{figure*}
\includegraphics[width=6.5in]{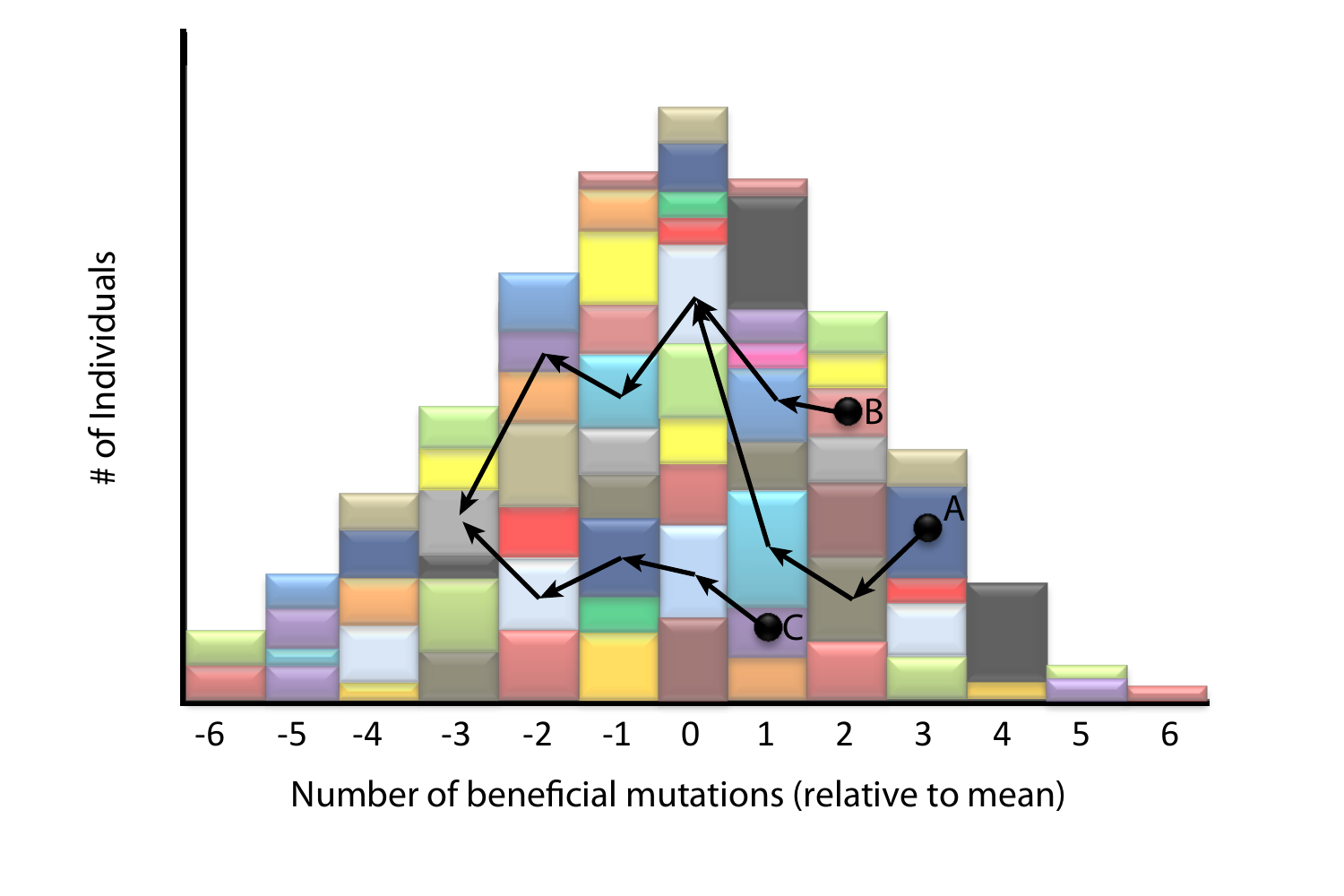}
\caption{Schematic of the fitness-class coalescent process.  The distribution of fitnesses within the population is shown (here for a case where the nose is ahead of the mean by $q = 6$ beneficial mutations).  Clonal lineages founded by individual beneficial mutations are shown in different colors within each fitness class.  Three individuals ($A$, $B$, and $C$) were sampled from the population, from classes $k=3$, $k=2$, and $k=1$ respectively.  The ancestors of individuals $A$ and $B$ descended from individuals in the silver lineage in fitness class $k=0$, and this individual shared a common ancestor with individual $C$ in the gray lineage in class $k = -3$.  Individuals $A$ and $B$ differ by $5$ beneficial mutations, while individual $C$ differs by $7$ beneficial mutations from the common ancestor of $B$ and $C$.  Individuals $A$ and $B$ coalesce when the silver lineage in class $k=0$ was originally created which occurred when this class was at the nose of the fitness distribution, $T_{AB} = 6 \bar \tau$ generations ago.  Individuals $A$, $B$, and $C$ last shared a common ancestor when the gray lineage in class $k = -3$ was originally created when this class was at the nose of the fitness distribution, $T_{MRCA} = 9 \bar \tau$ generations ago.  \label{fig4} }
\end{figure*}

We now describe the probability of a genealogy relating a sample of several individuals.  Imagine, for simplicity, that we sampled two individuals that both happened to be in the same fitness class, $k$.  If these individuals were from the same clonal lineage within that class, then they are genetically identical at all the $B$ positively selected sites.  We say they coalesced in class $k$  and did so when this class was at the nose of the fitness distribution, approximately $(q-k) \bar \tau$ generations in the past.  If these individuals were not from the same clonal lineage within the class, then they both descended from individuals, in what is now fitness class $k-1$, that got distinct beneficial mutations.  If the individuals in which these mutations occurred are from the same clonal lineage within class $k-1$, we say the sampled individuals coalesced in class $k-1$.  If so, they differ at two of the $B$ positively selected sites, and coalesced when class $k-1$ was at the nose of the fitness distribution, approximately $[q-(k-1)] \bar \tau$ generations ago.  If not, they descended from individuals, in what is now fitness class $k-2$, that got distinct beneficial mutations, and so on.  We can apply similar logic to larger samples or when the individuals were sampled from different fitness classes.  We illustrate this fitness-class coalescent process in \fig{fig4}.

We note that the probability a sample of individuals comes from the same clonal lineage is the same in each fitness class, since the clonal structure of the class was always determined when that class was at the nose of the distribution (nevertheless, conditional on some individuals coalescing in a class, the probability of additional coalescence events is substantially altered; see below).  In addition, the coalescence probabilities do not depend on when the mutations occurred in the ancestral lineages of each sampled individual, since all clonal lineages were founded when a class was at the nose of the fitness distribution.  These are major simplifications compared to the case of purifying selection, where the relative timings of mutations and the differences in clonal structure in different classes are important complications \citep{allelebased, negselcoalescent}.

To use the fitness-class coalescent approach to calculate the probability of a given genealogical relationship among a sample of individuals from the population, it only remains to calculate the probabilities that arbitrary subsets of these individuals coalesced within each fitness class.  In the next section, we use the above described  clonal structure  to compute these fitness-class coalescence probabilities.

\subsection{Fitness-class coalescence probabilities}

We begin our calculation of the fitness-class coalescence probabilities by considering the probability that $H$ individuals coalesce to $1$ in a given class.  We call this probability $\ph$.  This coalescence event will occur if and only if all $H$ of these individuals are members of the same clonal lineage.  The probability an individual is sampled from a clone of size $\nu$ is $\nu/\sigma$, so summing over all possible clones we have \eon \ph = \left\langle \sum_{i=1}^B \frac{\nu_i^H}{\sigma^H} \right\rangle \eoff with $\sigma\equiv \sum_i \nu_i$.  In Appendix A we use the expression for distribution of $\nu$ from \eq{two}, and take the $B \to \infty$ limit, to find \eon \ph = \frac{\Gamma(H - \alpha)}{\Gamma(H) \Gamma(1-\alpha)} . \label{ph} \eoff

We can use a similar approach to calculate the probabilities of more complicated coalescence configurations.  Consider the general situation where $H$ individuals coalesce into $K$ in a given fitness class, with $h_1$ individuals coalescing into lineage $1$, $h_2$ individuals coalescing into lineage $2$, and so on, up to $h_K$ individuals coalescing into lineage $K$ (note that $\sum_{j=1}^K h_j = H$).  In Appendix A, we show that this probability, $C_{H, K, \{h_j\}}$, is given by \eon C_{H, K, \{h_j\}} = \frac{H \alpha^{K-1}}{K} \prod_{j=1}^K \frac{\Gamma(h_j - \alpha)}{\Gamma(h_j +1) \label{pchk} \Gamma (1-\alpha)} . \eoff

In order to compute any quantity that depends on genealogical topologies, it will be important to know not just that $H$ individuals coalesced into $K$ lineages, but that they did so in a specific configuration $\{ h_j \}$. For example, if we have four individuals coalescing into two, this could occur by three of them coalescing into one and the other lineage not coalescing, or alternatively by two pairwise coalescence events.  These different topologies will affect some aspects of molecular evolution such as the polymorphism frequency distribution.  To compute these quantities, we must work with the full coalescence probabilities in \eq{pchk}.

However, the specific coalescence configurations do not affect non-topology-related quantities such as the total branch length, time to most recent ancestor, or any statistics that depend on these quantities (e.g. the total number of segregating sites $S_n$).  To compute the statistics of these aspects of genealogies, we only need to know $H$ and $K$. Thus it will be useful to sum the probabilities of all possible configurations $\{ h_j \}$ that lead to a particular $K$.  We call this total probability of $H$ individuals coalescing to $K$ lineages $\dcoal{H}{K}$. We have \eon \dcoal{H}{K} = \sum_{\{ h_j \}} C_{H, K, \{h_j\}} , \eoff  where the sum over the $\{ h_j \}$ is constrained to values such that $\sum_{j=1}^K h_j = H$.

To compute $\dcoal{H}{K}$, we first make the definition \eon f(H,K) = \sum_{\{ h_j \}} \prod_{j=1}^K \frac{\Gamma(h_j - \alpha)}{\Gamma(h_j +1) \Gamma (1-\alpha)} , \eoff and note that \eon \dcoal{H}{K} = \frac{H}{K \alpha} f(H,K). \label{fourteen} \eoff We can compute $f(H,K)$ using a simple contour integral, \eon f(H,K) = \frac{1}{2 \pi i} \int \frac{dz}{z^{H+1}} \left[ 1 - (1-z)^\alpha \right]^K, \eoff where the integral is taken circling the origin.  We can alternatively the generating function for $f(H,K)$, \eon R_f(z) \equiv \sum_{H=0}^\infty f(H,K) z^H. \eoff In Appendix A, we show that \eon R_f(z) = \left[ 1 - (1-z)^\alpha \right]^K. \eoff  We can now compute $f(H,K)$ for arbitrary $H$ and $K$ by noting that \eon f(H,K) = \frac{1}{H!}\frac{d^H}{dz^H} R_f(z) |_{z=0}, \eoff  and substitute this into \eq{fourteen} to compute $\dcoal{H}{K}$.  To give a few examples, we find \eaon \dcoal{2}{1} & = & \frac{1}{q} \\ \dcoal{3}{1} & = & \frac{1}{2q} \left( 1 + \frac{1}{q} \right) \\ \dcoal{3}{2} & = & \frac{3}{2q} \left( 1 - \frac{1}{q} \right). \eaoff  Taking more derivatives, we can easily make a table of $f(H,K)$ and evaluate any arbitrary $\dcoal{H}{K}$.  We note that in the large $H$ limit, one can directly obtain $f(H,K)$ using saddle point evaluation of the contour integral defined above.

Note that the case of rapid adaptation, for which clonal interference is pervasive, corresponds to the case where $q$ is reasonably large (conversely $q=1$ corresponds to sequential selective sweeps, and our analysis does not apply in this limit).  In the large-$q$ regime, $\dcoal{2}{1}$ is small.  In neutral coalescent theory, the probability of a three-way coalescence event would then be even smaller: $\dcoal{3}{1} \sim \dcoal{2}{1}^2$.  However, this is not the case here: the probability three lineages coalesce is of the same order as the probability two lineages coalesce, $\dcoal{3}{1} \sim \dcoal{2}{1}$, so ``multiple-merger'' coalescence events are not uncommon.  This is a signature of the fact that occasionally a fitness class is dominated by a single large clone, as described above.  When this happens, that clone dominates the structure of genealogies, as any ancestral lineages we trace through the fitness distribution are very likely to have originated from this single large lineage, and hence will coalesce within this fitness class.  Although these anomalously large clones are rare, they are sufficiently common that they are responsible for a significant fraction of the total coalescence events, and they are responsible for tendency of genealogies to take on a more ``star-like'' shape.

\section{Genealogies and Patterns of Genetic Variation}

From the results above for the probabilities of all possible coalescence events in each fitness class, we can calculate the probability of any genealogy relating an arbitrary set of sampled individuals.  From these genealogies, we can  in turn calculate the probability distribution of any statistic describing the expected patterns of genetic diversity in the sample.

We begin by neglecting neutral mutations and calculating the structure of genealogies in ``fitness-class'' space.  That is, we consider individuals sampled from some set of fitness classes.  We trace their ancestries backwards in time as they ``advance'' from one fitness class to the next, via mutational events, and calculate the probability that they coalesce in a particular set of earlier-established classes.  Since each step in the fitness-class coalescent tree corresponds to a beneficial mutation, this immediately gives us the  pattern of genetic diversity at the positively selected sites.  We  later consider how these ``fitness-class'' genealogies correspond to genealogies in real time, and use this to derive the expected patterns of diversity at linked neutral sites.

\subsection{The distribution of heterozygosity at positively selected sites}

We first describe the simplest possible case, a sample of two individuals.  If we sample two individuals at random from the population, the first comes from class $k_1$ and the second from class $k_2$ with probability $\h_{k_1} \h_{k_2}$.  If these two individuals coalesce in class $\ell$, their total pairwise heterozygosity at positively selected sites, $\pi_b$, will be $(k_1-\ell)+(k_2-\ell) = k_1 + k_2 - 2 \ell$.

We can now calculate the average $\pi_b$ given $k_1$ and $k_2$ by noting that \eon \ev{\pib{k_1}{k_2}} = |k_2 - k_1| + \ev{\pib{k}{k}}. \eoff By conditioning on whether two individuals sampled from class $k$ coalesce within that class (in which case they have $\pi_b=0$), we have \eon \ev{\pib{k}{k}} = 0 \dcoal{2}{1} + (1-\dcoal{2}{1}) \left[ \ev{\pib{k}{k}} + 2 \right], \eoff which implies \eon \ev{\pib{k}{k}} = \frac{2 (1-\dcoal{2}{1})}{\dcoal{2}{1}}.  \eoff  Plugging this into the above, we find \eon \ev{\pib{k_1}{k_2}} = |k_2 - k_1| + \frac{2(1-\dcoal{2}{1})}{\dcoal{2}{1}}. \eoff  We can now average this over $k_1$ and $k_2$ to find the overall average.  Since $k_1$ and $k_2$ are approximately normally distributed with variance $1/(s\bar \tau)$, their average absolute difference is $\sqrt{4/(s\bar{\tau}\pi)}$.  Thus we have \eon \ev{\pib{k_1}{k_2}} = \sqrt{\frac{4}{\pi s\bar \tau}} + \frac{2(1-\dcoal{2}{1})}{\dcoal{2}{1}}. \eoff  Note that for large $q$, the second term (corresponding to heterozygosity between individuals sampled from the same class) is approximately $2q$, while the first term is approximately $\sqrt{4q/\pi\log(s/U_b)}$, which is smaller by a factor of $1/\sqrt{2\pi \log(Ns)}$. This is because most individuals are much closer to the mean than to the nose, so that $|k_1-k_2|\ll q$.   In other words, a rough but very simple approximation is to assume that all individuals are sampled from the mean fitness class.

\begin{figure*}
\includegraphics[width=6.5in]{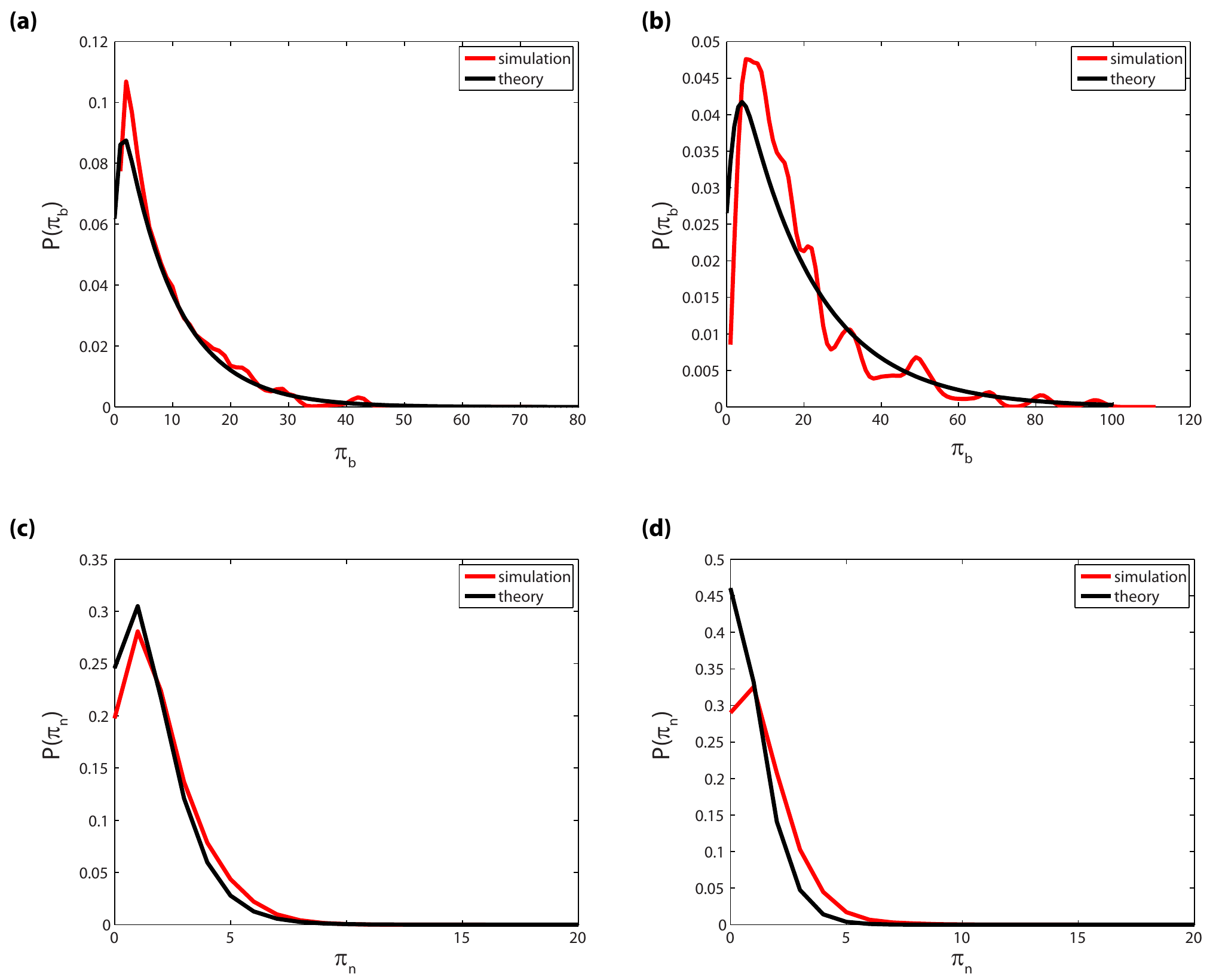}
\caption{The distribution of pairwise heterozygosity.  (\textbf{a}) Comparison of our theoretical predictions for the distribution of pairwise heterozygosity at positively selected sites, $\pi_b$ with the results of forward-time Wright-Fisher simulations, for $N = 10^7$, $s = 10^{-2}$, and $U_b = 10^{-4}$.  Simulation results are an average over $56$ independent runs, with $10^6$ pairs of individuals sampled from each run. (\textbf{b}) Pairwise heterozygosity at positively selected sites for $N = 10^7$, $s = 10^{-2}$, and $U_b = 10^{-3}$.  (\textbf{c}) Comparison of our theoretical predictions for the distribution of pairwise heterozygosity at linked neutral sites, $\pi_n$ with the results of forward-time Wright-Fisher simulations, for $N = 10^7$, $s = 10^{-2}$, $U_b = 10^{-4}$, and $U_n = 10^{-3}$.  (\textbf{d}) Pairwise heterozygosity at linked neutral for $N = 10^7$, $s = 10^{-2}$, $U_b = 10^{-3}$, and $U_n = 10^{-3}$.  \label{fig5} }
\end{figure*}

We can use a similar approach to compute the full probability distribution of $\pi_b$. We have \eon P(\pib{k}{k} = \gamma) = \dcoal{2}{1} \delta_{\gamma, 0} + (1 - \dcoal{2}{1}) P(\pib{k}{k} = \gamma-2), \eoff which implies that \eon P(\pib{k}{k} = \gamma) = \left\{ \begin{array}{ll} \dcoal{2}{1} (1-\dcoal{2}{1})^{\gamma/2} & \textrm{for } \gamma \textrm{ even} \\ 0 & \textrm{for } \gamma \textrm{ odd} \end{array} \right. . \eoff  We can then write the more general result \eon P(\pib{k_1}{k_2} = \gamma) = \dcoal{2}{1} \delta_{\gamma, k_1 - k_2} + (1-\dcoal{2}{1}) P(\pib{k_1}{k_2} = \gamma - 2), \eoff from which we find \begin{widetext} \eon P(\pib{k_1}{k_2} = \gamma) = \left\{ \begin{array}{ll} \dcoal{2}{1} (1-\dcoal{2}{1})^{\frac{\gamma-(k_1-k_2)}{2}} & \textrm{for } \frac{\gamma-(k_1-k_2)}{2} \textrm{ even and } \gamma \geq k_1-k_2 \\ 0 & \textrm{otherwise} \end{array} \right. . \eoff
\end{widetext} If desired, we can now average these results over the distributions of $k_1$ and $k_2$ to get the unconditional distribution of $\pi_b$.  In \fig{fig5}a and \fig{fig5}b, we illustrate these theoretical predictions for the overall distribution of pairwise heterozygosity with the results of full forward-time Wright-Fisher simulations, for two representative parameter combinations.  We see that the distribution of heterozygosity has a nonzero peak, and that the agreement with simulations is generally good.

We emphasize that our results for $P(\pi_b)$ describe the \emph{ensemble} distribution of heterozygosity.  That is, if we picked a single pair of individuals from each of many {\it independent populations}, this is the distribution of $\pi_b$ one would expect to see.  It is \emph{not} the  population distribution: if we were to pick many pairs of individuals from the same population, the $\pi_b$ of these pairs would  not be independent because much of the coalescence within individual populations occurs in rare classes that are dominated by a single lineage for which $\dcoal{2}{1}$ is much higher than its average value.  Thus if we measured the average $\pi_b$ within each population by taking many samples from it, the distribution of this $\bar{\pi_b}$ across populations would be different from the distribution computed above.  In order to understand these within-population correlations, we now consider the genealogies of larger samples.

\subsection{Statistics in larger samples}

We can compute the average and distribution of statistics describing larger samples in an analogous fashion to the pair samples.  For example, consider the total number of segregating positively selected sites among a sample of $3$ individuals, which we call $S_{3b}$.  These three individuals are sampled (in order) from classes $k_1$, $k_2$, and $k_3$ respectively with probability $\h_{k_1} \h_{k_2} \h_{k_3}$.  For three individuals sampled from the same fitness class $k$, by conditioning on the coalescence possibilities within class $k$ we find that the average total number of segregating positively selected sites is \eon \ev{S_{kkk}} = 0 \dcoal{3}{1} + \dcoal{3}{2} \left[ 2 + \ev{\pib{k}{k}} \right] + \dcoal{3}{3} \left[ 3 + ev{S_{kkk}} \right]. \eoff  Solving this for $\ev{S_{kkk}}$, we find \eon \ev{S_{kkk}} = \frac{2 \dcoal{3}{2}/\dcoal{2}{1} + 3 \dcoal{3}{3}}{\dcoal{3}{1} + \dcoal{3}{2}}. \eoff  More generally we have \eaon \ev{S_{k_1 k_2 k_2}} & = & (1-\dcoal{2}{1})^{k_2-k_1} \left[ 2 (k_2 - k_1) + \ev{S_{kkk}} \right] \\ & & \nonumber + \sum_{i=0}^{k_2-k_1-1} \dcoal{2}{1} (1-\dcoal{2}{1})^i \left[ k_2 - k_1 + \pib{k}{k} + i \right], \eaoff and even more generally we have \eon \ev{S_{k_1k_2k_3}} = k_3 - k_2 + \ev{S_{k_1 k_2 k_2}}. \eoff  If desired, we can average these over the distribution of $k_1$, $k_2$, and $k_3$ using the properties of differences of Gaussian random variables, as above.  Alternatively, as in samples of size two, in large populations we can make the rough approximation that all sampled individuals come from the mean fitness class. Analogous calculations can be used to find the average number of segregating positively selected sites in still larger samples.

\begin{figure}
\includegraphics[width=3.2in]{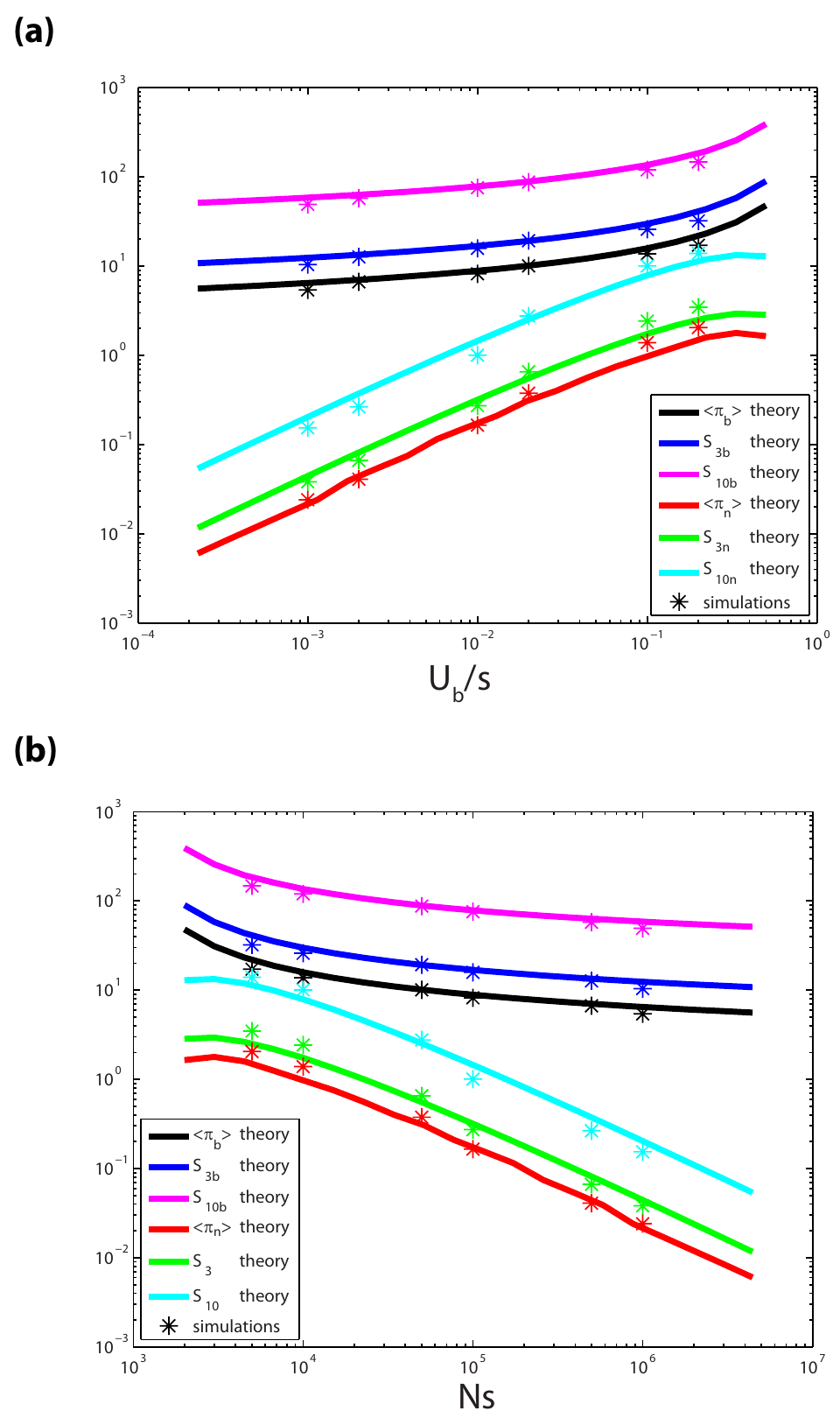}
\caption{Comparisons between theoretical predictions (from coalescent simulations) and forward-time Wright-Fisher simulations for the average pairwise heterozygosity and total number of segregating sites in samples of size $3$ and $10$ at positively selected sites and at linked neutral sites, (\textbf{a}) as a function of $U_b/s$ and (\textbf{b}) as a function of $Ns$.  In both panels, $N = 10^7$ and $U_b = 10^{-4}$ while $s$ is varied.  Forward-time Wright-Fisher simulation data represents an average over $56$ forward simulation runs, with $10^6$ pairs of individuals sampled from each run.  Theoretical predictions generated using backwards-time coalescent simulations represent the average of $3 \times 10^6$ independently simulated pairs of individuals.  Note that both (\textbf{a}) and (\textbf{b}) show the same data, plotted as a function of different parameters.  \label{fig6} }
\end{figure}

In \fig{fig6} we illustrate some of these predictions (in practice samples are generated from coalescent simulations; see below) for samples of size $2$, $3$, and $10$, and compare these to the results of forward-time Wright-Fisher simulations.  We note that the agreement is generally good.

We can apply similar thinking to describe the distribution of the total number of segregating selected sites.  First consider this distribution for a sample of size $3$, all of which happen to be sampled from the same fitness class $k$, $S_{kkk}$.  We have \eaon P(S_{kkk} = \gamma) & = & \dcoal{3}{1} \delta_{\gamma, 0} + \dcoal{3}{2} P(\pib{k}{k} = \gamma-2) \nonumber \\ & & + \dcoal{3}{3} P(S_{kkk} = \gamma - 3). \eaoff  We can multiply by $z^\gamma$ and sum over $\gamma$ to pass to generating functions, $U_3(z) \equiv \sum z^\gamma P(S_{kkk} = \gamma)$.  This yields \eon U_3(z) = \dcoal{3}{1} + \dcoal{3}{2} z^2 U_2(z) + \dcoal{3}{3} z^3 U_3(z), \eoff which we can solve to find \eon U_3(z) = \frac{\dcoal{3}{1} + z^2 \dcoal{3}{2} U_2(z)}{1 - \dcoal{3}{3} z^3}, \eoff where we have introduced the obvious notation.

More generally, we have that the total number of segregating sites among a sample of $H$ individuals all chosen from the \emph{same} fitness class $k$, which we will call $S_H$, has the distribution \eaon P(S_H = \gamma) & = & \dcoal{H}{1} \delta_{\gamma, 0} + \dcoal{H}{2} P(S_2 = \gamma - 2) + \\ & & \nonumber \dcoal{H}{3} P(S_3 = \gamma - 3) + \ldots \dcoal{H}{H} P(S_H = \gamma - H). \eaoff  We can again pass to generating functions, giving \eon U_H(z) = \dcoal{H}{1} + \dcoal{H}{2} z^2 U_2(z) + \dcoal{H}{3} z^3 U_3 + \ldots, \eoff which we can easily solve to give \eon U_H(z) = \frac{\dcoal{H}{1} + \sum_{\ell=2}^{H-1} z^\ell \dcoal{H}{\ell} U_{\ell}(z)}{1 - \dcoal{H}{H} z^H}. \eoff

It still remains to consider the distribution of the total number of segregating selected sites among $H$ individuals chosen at random from arbitrary fitness classes.  The general case becomes quite unwieldy to compute analytically, because we must average over all fitness classes in which internal coalescence events can occur.  Computing these averages for the case of a sample of size three, we find that the generating function for the distribution of the total number of segregating positively selected sites among a sample of three individuals sampled from classes $k_1$, $k_2$, and $k_3$ is given by \eaon W_3(z|k_1, k_2, k_3) & = & \frac{z^{k_1-k_3} U_2(z) \dcoal{2}{1} \left[ 1 - \left( z \dcoal{2}{2} \right)^{k_1-k_2} \right]}{1 - \dcoal{2}{2} z} \nonumber \\ & & + \dcoal{2}{2}^{k_1-k_2} U_3(z). \eaoff  Note that these distributions are all for samples each taken from an independently evolved population, rather than found from averaging many samples from each population and then finding the distribution of this across populations.

Analogous expressions can be computed for larger samples, but these involve ever more complex combinatorics.  One may also  wish to compute other statistics describing genetic variation in larger samples, such as the allele frequency spectrum.  While in principle it is possible to calculate analytic expressions for any such statistic using methods similar to those described above, in practice it is easier to use our fitness-class coalescent probabilities to implement coalescent simulations, and then use these simulations to compute any quantity of interest. We describe these coalescent simulations in a later section.  Alternatively, for large populations we can make use of the rough approximation that all individuals are always sampled from the mean fitness class; we explore some consequences of this approximation further in a later section below.

\subsection{Time in generations and neutral diversity}

Thus far we have focused on the fitness-class structure of genealogies and the genetic variation at positively selected sites.  We now describe the correspondence between our ``fitness-class coalescent'' genealogy and the genealogy as measured in actual generations.  Fortunately, this correspondence is extremely simple: each clonal lineage was originally created by mutations when that fitness class was at the nose of the fitness distribution.  Thus if we define the current mean fitness to be class $k=0$, the current nose class will be at approximately $k=q$, and some arbitrary class $k$ will have been created at the nose approximately $(q-k) \bar \tau$ generations ago.  Although there is some variation in each establishment time, we neglect this variation throughout our analysis here, since it is small compared to the variation between coalescence times within clones in different classes.  As we will see below, this approximation holds well in comparison to simulations in the parameter regimes we consider.  This makes the correspondence between real times and steptimes much simpler here than in our previous analysis of purifying selection, where the variation in real times, even given a specific fitness-class coalescent genealogy, was substantial \citep{negselcoalescent}.

The simple approximation of neglecting the variations in time of establishment of the fitness classes allows us to make a straightforward deterministic correspondence between the fitness-class coalescent genealogy and the coalescence times.  We can then compute the expected patterns of genetic diversity at linked neutral sites: the number of neutral mutations on a genealogical branch of length $T$ generations is Poisson distributed with mean $U_n T$.  From this we can compute the distribution of statistics describing neutral variation (e.g. the neutral heterozygosity $\pi_n$ or total number of neutral segregating sites in a sample $S_n$) from the corresponding statistics describing the variation at the positively selected sites.  We illustrate these theoretical predictions for the distribution of neutral heterozygosity $\pi_n$ in \fig{fig5}c and \fig{fig5}d, and compare these predictions to the results of full forward-time Wright-Fisher simulations.  In \fig{fig6} we also show our predictions (generated using the coalescent simulations described above) for the mean number of segregating neutral sites in samples of size $2$, $3$, and $10$, compared to the results of forward-time Wright-Fisher simulations.  We note that the agreement is good across the parameter regime we consider, though there are some systematic deviations for smaller values of $U_b/s$ where our approximations are expected to be less accurate.

\subsection{Time to the  Most Recent Common Ancestor}

Thus far we have considered the coalescence events at each mutational step separately: this is necessary to describe the full structure of genealogies.  However, another important quantity of interest is the time to the most recent common ancestor --- i.e. the coalescence time of the entire sample.  We begin by considering this time measured in mutational steps, and then describe how this relates to the coalescence time measured in generations.

We can derive relatively simple expressions for the number of mutational steps to coalescence of an entire sample by directly calculating the probability of coalescence events over several steps at once.  To do so, we note that since the dynamics at each mutational step are identical, the generating function of the number of individuals descended from a mutation at site $i$ that occurred $\ell$ mutational steps ago, $\nu_i^{(\ell)}$, is given by \eon G_i^{(\ell)}(z) = \ev{e^{-z} \nu_i^{(\ell)}} = \exp \left[ -\frac{1}{B} z^{\eta_\ell} \right], \eoff where we have defined \eon \eta_\ell \equiv \alpha^\ell = \left( 1 - 1/q \right)^\ell .  \eoff

From this expression, we can immediately compute the distribution of the number of mutational steps to coalescence of $H$ individuals sampled from the same fitness class, $\tmrca (H)$.  The cumulative distribution of $\tmrca$ is given by \eon F(H,\ell) \equiv \mathrm{Prob} \left[ \tmrca(H) \leq \ell \right] \approx \sum_{i=1}^B \left( \frac{\nu_i^{(\ell)}}{\sum_{i=1}^B \nu_i^{(\ell)}} \right)^H. \eoff  We can compute $\ev{F(H,\ell)}$ using identical methods to those used to calculate the fitness-class coalescence probabilities above, and find \eon \ev{F(H,\ell)} = \frac{\Gamma(H-\eta_\ell)}{\Gamma(H) \Gamma(1-\eta_\ell)}. \eoff  From this, we find \eon \ev{\tmrca(H)} = \sum_{\ell=0}^\infty \left( 1 - \ev{F(H,\ell)} \right). \eoff   Note that we could alternatively obtain expressions for $\tmrca(H)$ more directly from the fitness-class coalescence probabilities in a single step, by conditioning on the coalescence events that can happen in the first step in a similar way to that we used to compute $\ev{\pi_b}$ and $\ev{S_{3b}}$.

In the large-$q$ limit, the ratios of these coalescence times (measured in mutational steps) in samples of different sizes are independent of $q$: \eon \frac{\ev{\tmrca(3)}}{\ev{\tmrca(2)}} = \frac{5}{4}, \quad \frac{\ev{\tmrca (4) }}{\ev{\tmrca (2) }} = \frac{25}{18}, \quad \frac{\ev{\tmrca(5)}}{\ev{\tmrca(2)}} = \frac{427}{288}.  \eoff These ratios are identical to those given by the Bolthausen-Sznitman coalescent \citep{bolthausensznitman98}, which has recently been shown to describe a number of other very different models of selection \citep{brunetderrida12b}.  We return to this point in the Discussion.  For large $H$ we find \eon \frac{\ev{\tmrca(H)}}{\ev{\tmrca(2)}} \to \log \log H + \mathcal{O}(1). \eoff These results suggest that there is a  $q$-independent limiting process: we discuss this briefly below.  We also note that the distribution of times to coalescence for large $H$ is quite different than in the neutral case --- the between-populations variation in $ \tmrca(H)/\ev{\tmrca(2)}$ is only of order unity, compared to its mean of $\log\log H$. In contrast, for the neutral coalescent, the time to last common ancestor of the whole population has mean  of  $2\ev{\tmrca(2)}$  and random variations of the same order.

As with other aspects of genealogical structures, it is straightforward to convert these expressions for the coalescence times measured in mutational steps to the time in generations to the most recent common ancestor of a sample, $T_{MRCA}(H)$.  Specifically, $\tmrca = \ell$ corresponds to the case where the most recent ancestor occurs $\ell$ mutational steps ago, so if the sampled individuals were from class $k$ the time to the most recent common ancestor is $\left[ q - (k-\ell) \right] \bar \tau$ generations.  We note that for a sample of two this implies that the nose-to-mean time $\tsw$ is the characteristic time scale of the coalescent, as claimed above.

Thus far we have considered the most recent common ancestor of $H$ individuals all sampled from the same fitness class $k$.  However, in general we will typically sample individuals from a variety of different classes.  In this case, we must sum over all possible internal coalescence events, until we reach a state where all remaining ancestral lineages are together in the same fitness class.  This quickly becomes unwieldy in larger samples.  In practice, it is easier to compute times to the most common recent ancestor in these cases using coalescent simulations based on our fitness-class coalescent approach, which we describe below.

As with other statistics described above, however, there is a simple approximation which is asymptotically correct for large populations: we can simply assume that all individuals are sampled from the mean fitness class.  This approximation relies on the fact that most individuals sampled randomly from the population will have fitnesses close to the mean: within of order $\sqrt{v}$ of it.   Thus the time differences between their establishments will typically be substantially smaller than the nose-to-mean time, $\tsw$.   As this is the time scale on which typical coalescent events take place, treating all the individuals as if they were in the dominant fitness class is a reasonable rough approximation. In this approximation, the results for the times to most common ancestor for samples of $H$ can be simply obtained from the single-fitness class results above. We find: \eon \ev{T_{MRCA}(2)}\approx 2\tsw , \eoff and in larger samples we have \eon \frac{\ev{T_{MRCA}(3)}}{\ev{T_{MRCA}(2)}} = \frac{9}{8}, \nonumber \eoff \eon \frac{\ev{T_{MRCA} (4)}}{\ev{T_{MRCA} (2)}} = \frac{43}{36}, \eoff \eon \nonumber \frac{\ev{T_{MRCA}(5)}}{\ev{T_{MRCA}(2)}} = \frac{715}{576}.  \eoff  We note however that the dominant-fitness-class approximation is valid only in the limit that the lead of the population, $qs$, is much larger than the standard deviation of the fitness distribution, $\sqrt{v}$.  As this ratio is $\sqrt{2\log(Ns)}$, in practice it never becomes very large.

\subsection{The Frequency of Individual Mutations}

An alternative way to compute many of the coalescent properties is to consider the fraction of the population with a particular mutation, which is closely related to the site frequency spectrum.  The frequency of a given mutation at a particular site is determined by when that mutation occurred relative to others in its fitness class.  In addition, its frequency at later times is determined by whether or not later mutations occur in its genetic background at each subsequent mutational step.  Consider a mutation that occurred $\ell$ steps in the past, and define $f\equiv \frac{\nu}{\sigma}$ to be the fraction of the current nose class that its descendants constitute.  The probability density of $f$ is \eon \rho_\ell(f)df= \frac{df}{B} \frac{1}{\Gamma(\eta_\ell)\Gamma(1-\eta_\ell)f^{1+\eta_\ell}(1-f)^{1-\eta_\ell}}\ , \eoff where as before we have defined $\eta_\ell = (1 - 1/q)^\ell$.   Coalescent properties depend on averages of $f^H$.   Summing over all $B$ sites and using the standard integrals of powers of $f$ and $1-f$ expressed in terms of gamma functions, we obtain immediately the result we had found above: $\ev{F(H,\ell)} = \frac{\Gamma(H-\eta_\ell)}{\Gamma(H) \Gamma(1-\eta_\ell)}$.

More generally, one can consider how the frequency of a mutation changes in time due to successive mutations in its lineage.  If a given mutation has frequency $g$ at one time, then a time $\ell\bar{\tau}$ later (after $\ell$ further beneficial mutations have occurred) the probability density of its frequency will be: \begin{widetext} \eon \rho_\ell(f|g)df= df\, \frac{ \frac{g (1 - g)}{\Gamma( \eta_\ell) \Gamma(1 - \eta_\ell)f(1 - f)}}{(1-g)^2 \left(\frac{f}{1-f}\right)^{\eta_\ell} + g^2\left(\frac{1-f}{f}\right)^{\eta_\ell} +2g(1-g)\cos(\pi{\eta_\ell})} \ . \eoff \end{widetext} From this, quantities such as the variance of the probability of $H$ individuals coalescing $\ell$ steps in the past and hence the variances in the coalescent times of $H$ individuals can be computed.

In the limit of large $q$, the exponent $\eta$ that parameterizes the time difference, $t=\ell\bar{\tau}$, is simply $\eta\approx e^{-t/\tsw}$.  This is independent of $q$: only the ``nose-to-mean" time that it takes for the new mutants to dominate the population matters.  In this limit, a single mutational step occurs in a time that is a very small fraction, $\epsilon=1/q$, of the nose-to-mean time $\tsw$.  The conditional probability of going from $g$ to $f$ in this step is \eon \rho_\ell(f|g)df\approx \frac{g(1-g)\epsilon \ df}{(f-g)^2+\pi^2\epsilon^2[g(1-g)]^2} \ . \label{rhodelta}\eoff Eq.~\ref{rhodelta} is an approximate delta-function in $f-g$, as one would expect in the limit of a small time step.  But it also corresponds to a probability per unit time of a jump from $g$ to $f$ of $\frac{1}{\tsw} df g(1-g)/(f-g)^2$. Specifically it describes the genetic background either containing the mutation (frequency $g$) or not containing the mutation (frequency $1-g$) increasing in size by a factor between $1+h$ and $1+h+dh$ with rate $\frac{1}{\tsw} dh/h^2 $ (with $\epsilon$ providing a small $h$ cutoff).  This corresponds  to a continuous time birth process in a sub-population of (large) size $n$ with rate per individual to give birth to $k$ offspring, $\frac{1}{\tsw} \frac{1}{k^2}$.  These considerations provide an alternative way to compute coalescent statistics.

\subsection{Coalescent Simulations}

We can use the fitness-class coalescence probabilities in \eq{pchk} to implement an algorithm for coalescent simulations along the lines of \citet{gordocharlesworth02}, using the structured coalescent framework of \citet{hudsonkaplan94}.  Specifically, to describe the diversity in a sample of $n$ individuals, we first randomly sample their fitness classes independently from the distribution $\h_k$.  We then start with the individual in the most-fit class, and trace back its ancestry as it steps through successive classes within the fitness distribution.  When that individual enters a class with other individuals, we use \eq{pchk} to determine the probabilities of all possible coalescence events in that class. We then continue to trace back the ancestry of the sample further through the distribution, allowing for coalescence events at each step according to the appropriate probabilities.  We continue this procedure until all individuals have coalesced.

This simple coalescent algorithm produces a fitness-class coalescent tree drawn from the appropriate probability distribution of genealogies.  We can then compute any statistic of interest describing this genealogy.  By repeating this algorithm, we can obtain the probability distribution of the statistic.  In practice this is a highly efficient procedure, since the coalescent simulations are extremely fast and the computational time required scales only with the size of the sample rather than the size of the population.

\subsection{Comparison to Simulations}

Our coalescent simulations represent an algorithmic implementation of our fitness-class coalescent, using all of the analytical expressions for the sampling and coalescence probabilities described above. Thus these coalescent simulations rely on all of the approximations underlying our method.  To test the validity of these approximations and the accuracy of our fitness-class coalescent method, we compared the predictions of these coalescent simulations to full forward-time Wright-Fisher simulations of our model.  These comparisons are illustrated in \fig{fig5} and \fig{fig6} and in Table \ref{table1}.

\begin{table}
\begin{tabular}[c]{|c|c|c|c|}
 \hline
  $u_b/ s$ &  $Ns$ &  $D_{10}$ theory& $D_{10}$  simulations  \\
    \hline
      0.2000   &   5000  &-3.3199  & -3.3378\\
        \hline
    0.1000  & 10000 & -3.3489  & -3.3569\\
      \hline
    0.0200 & 50000  &  -3.3533   &-3.3322\\
      \hline
    0.0100  & 100000  & -3.3571 &  -3.4188\\
      \hline
    0.0020  & 500000  & -3.3665   &-3.3024\\
      \hline
    0.0010   & 1000000 &-3.3717 &  -3.3670\\
   \hline
\end{tabular}
\caption{Comparisons between theoretical predictions (from coalescent simulations) and forward-time Wright-Fisher simulations for Tajima's D \citep{tajima89} in a sample of size $10$, $D_{10}$.  Here $U_b = 10^{-4}$ and $N = 10^7$ while $s$ is varied.  Theoretical predictions are obtained by sampling $10^7$  backward coalescent simulations.  Forward-time simulation results are an average over $56$ forward simulation runs, with $10^6$ samples of $n=2$ and $n=10$ individuals. \label{table1}}
\end{table}

Our Wright-Fisher simulations were implemented assuming a population of constant size $N$, in which each generation consisted of a mutation and a selection step.  In the mutation step, we independently choose the number of beneficial and neutral mutations within each extant genotype from the appropriate multinomial distribution.  Each new mutation was assigned a unique index and all unique genotypes were tracked.  In the selection step, we sample $N$ individuals with replacement from the previous generation, using a multinomial sampling weight adjusted for selective differences between individuals relative to the population mean fitness \citep{ewensbook}.

\section{Discussion}

We have developed a fitness-class coalescent method to calculate how positive selection on many linked sites alters the structure of genealogies.  This has allowed us to calculate how clonal interference shapes the patterns of genetic diversity in rapidly adapting populations.  Our approach moves away from the traditional method of calculating the structure of genealogies in real time.  Rather, we treat each mutational step from one fitness class to the next as an ``effective generation,'' and trace how a sample of individuals descended by mutations through these fitness classes.  In each ``effective generation'' we calculated the total probability of all possible coalescence events, \eq{pchk}.  This allows us to calculate the structure of genealogies in this ``fitness-class space,'' which directly corresponds to the genetic diversity at positively selected sites.  We then converted this fitness-class coalescent to the genealogy in real time in order to calculate the expected patterns of neutral diversity.

We have shown that we can use this approach to compute analytic expressions for the distributions of several simple statistics describing patterns of molecular evolution.  However, it is often easiest to compute expected patterns of variation using backwards-time coalescent simulations which explicitly implement the fitness-class coalescent algorithm using the distribution of the fraction of the population in each fitness class $\h_k$ and the coalescence probabilities in \eq{pchk} to simulate genealogies.  These coalescent simulations are extremely efficient, and in practice it is usually faster to run millions of these backwards-time simulations than it is to numerically evaluate the sums over fitness classes involved in the corresponding exact analytic expressions.  These coalescent simulations also have the advantage of being very similar in spirit to structured coalescent simulations that describe the effects of purifying selection (see e.g. \citet{gordocharlesworth02} and \citet{seger10}), so they can in principle be used for parameter estimation and inference in analogous ways.

Our analysis throughout this paper is very similar in spirit to the fitness-class coalescent method we previously used to describe how purifying selection at many linked sites alters the structure of genealogies and patterns of molecular evolution \citep{allelebased, negselcoalescent}.  However, there are two important technical differences.  First, in the case of purifying selection, fluctuations in the frequencies of each fitness class $\h_k$ due to genetic drift can be substantial in certain parameter regimes.  These fluctuations are particularly important near the nose of the distribution, where they can lead to effects such as Muller's ratchet.  Although individuals are unlikely to be sampled from this nose, they are very likely to coalesce there.  Neglecting these fluctuations was therefore an important approximation that substantially restricted the regime of validity of our analysis.  By contrast, in the case of positive selection, fluctuations in the sizes of each fitness class are negligible (except at the nose) across a broad range of relevant parameter values.  Furthermore, fluctuations at the nose are much less important for patterns of diversity than in the case of purifying selection, because individuals are unlikely to either be sampled there or to coalesce there.  This reflects a fundamental difference between the neutral and purifying selection processes and the rapid adaptation dynamics analyzed here.  For the former, genetic drift plays a key role in driving the fluctuations, while for the latter, genetic drift is almost irrelevant: the fluctuations are dominated by the stochasticity in the timings of the beneficial mutations that occur near the nose of the fitness distribution.

A second key simplification of our analysis of positive selection, compared to the purifying selection case, is that the clonal structure of each fitness class becomes effectively ``frozen'' once that class is no longer at the nose of the fitness distribution.  This means that coalescence probabilities are identical in all fitness classes which stands in contrast to the case of purifying selection, where the clonal structure within all classes is constantly changing.  This also avoids the need to carefully analyze the timing and order of mutation events in the history of a sample  and simplifies the mapping between our fitness-class coalescent genealogy and the genealogy measured in real time.

\begin{figure*}
\includegraphics[width=6.5in]{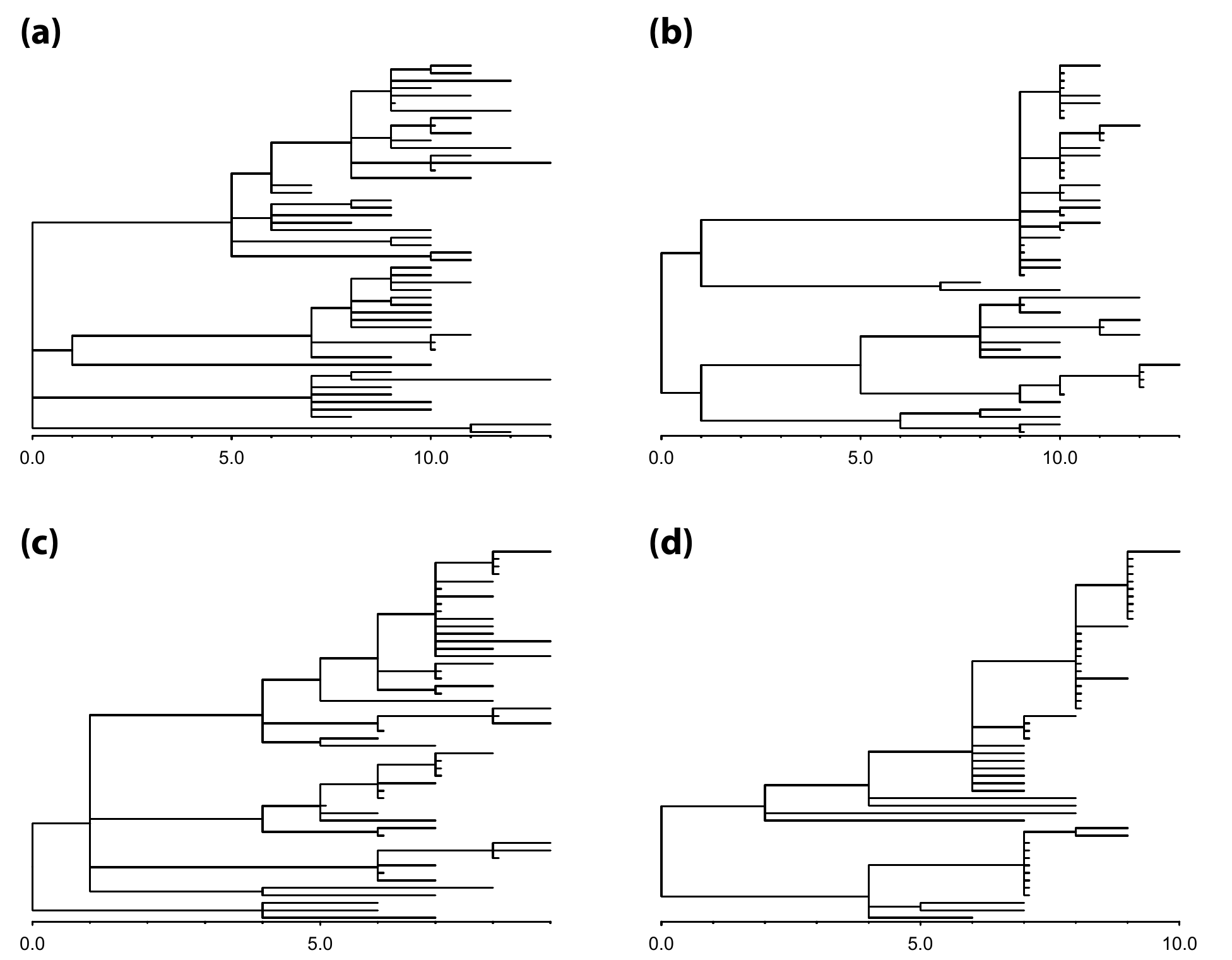}
\caption{Examples of fitness-class coalescent genealogies in samples of size $50$ from forward-time Wright-Fisher simulations.  The tips of each tree correspond to individuals sampled from the present.  Each tip is placed horizontally according to the fitness class from which that individual was sampled (classes are numbered according to the number of beneficial mutations relative to the most recent common ancestor of the sample).  Coalescence events are depicted according to the fitness class in which they occurred.  Each unit of time on the horizontal axis corresponds to one beneficial mutation, so that two individuals separated by a branch length of $\ell$ have $\pi_b = \ell$.  These fitness-class genealogies can be converted to genealogies  in real time by using our approximation that all coalescent events happen when the relevant class was at the nose of the fitness distribution. Note that the characteristic time for coalescence is the time it takes for $q$ successive beneficial mutations: this varies considerably with the parameters used. In all trees, $N = 10^7$ and $U_b = 10^{-4}$.  (\textbf{a}) An example of a genealogical tree for $s = 10^{-3}$.  (\textbf{b}) An example of a tree for $s = 5 \times 10^{-3}$.  (\textbf{c}) An example of a tree for $s = 10^{-2}$.  (\textbf{d}) An example of a tree for $s = 5 \times 10^{-2}$.  \label{fig7} }
\end{figure*}

Our results demonstrate how positive selection on many linked sites distorts the structure of genealogies away from neutral expectations.  We show several examples of these selected genealogies, for various different parameter values, in \fig{fig7}.  The most striking qualitative conclusion of our analysis is that multiple merger events, where several ancestral lineages coalesce into one in a single effective generation, occur with comparable probabilities to pairwise coalescence events.  We note that these events are multiple mergers within a single effective generation in our fitness-class coalescent, and hence are not actually multiple mergers within a single real generation.  However, these events happen very close together in real time compared to the other relevant timescales, so they will appear as effectively instantaneous.  This leads to a more ``starlike'' shape of genealogical trees.  This signature is characteristic of the action of positive selection; our analysis here illustrates how starlike we expect genealogies to be (and how many deeper coalescence events are preserved) given the interplay between interference and hitchhiking effects characteristic of this rapid adaptation regime.  It may prove useful in future work to analyze this specific situation in the context of more general models of the coalescent with multiple mergers \citep{pitman99}.

We note that the characteristic time scale of the coalescence is the ``nose-to-mean" time, $\tsw$, which is the time after which the collection of new mutants at the nose take to dominate the population.  In units of this time, trees for different values of $q$ become statistically similar for large $q$.  One striking feature, that occurs roughly once each $\tsw$, is the coalescence of a substantial fraction of all the (remaining) lineages at a single time step: this is caused by one new beneficial mutation occurring so much earlier than typical that its descendants represent a substantial fraction of the population in the nose.  Examples of this can be seen in Fig. 7.  Another perhaps-surprising feature of the genealogies in large samples is that some aspects are {\it less} variable from one population to another than neutral coalescent trees, while other aspects are more variable.  In the recent past, for times much shorter than the mean coalescence time of pairs of individuals, neutral coalescent trees, tend to be rather similar, while the multiple-coalescence events that characterize the positively selected genealogies cause larger variations between populations.   In contrast, the time to last common ancestor of large samples is broadly distributed for neutral trees but narrowly distributed (at least asymptotically) for positively selected trees.

Because individuals are unlikely to be sampled from near the nose of the distribution, the initial coalescence events in the history of the sample are typically in the bulk of the fitness distribution.  Since these coalescence events happened well in the past when these classes were at the nose of the distribution, the terminal branches in the genealogies of a sample are likely to be longer compared to internal branches than we would expect under neutrality.  In other words, recent branches of genealogies are longer relative to more ancient branches.  This effect is qualitatively similar to the situation in which effective population size declines as time recedes into the past: this has long been recognized as a general signature of the effects of both purifying and positive selection.  It leads to an excess of singleton mutations in the site frequency spectrum, and the negative values of Tajima's $D$ that we have observed.  However, clonal interference mitigates these effects relative to a hard selective sweep.

Our results also demonstrate that even when beneficial mutations are rare compared to neutral mutations, $U_b \ll U_n$, positively selected sites can still contribute a significant fraction of the total genetic variation observed in a population.  For example, in a sample of two individuals the total heterozygosity at positively selected sites will typically be several times $q$.  The typical neutral heterozygosity, on the other hand, will be of order $\pi_n \sim U_n \tsw$.  Thus even when $U_n \gg U_b$, $\pi_b$ will often be comparable to or even greater than $\pi_n$.  This is consistent with the general observation in microbial evolution experiments that a substantial fraction of observed mutations are beneficial \citep{gresham08, kaosherlock08, barrick09a, barrick09b}.  The fact that positively selected sites can be a significant fraction of the polymorphisms emphasizes the importance of understanding the patterns of diversity at these sites, which have distinct patterns compared to linked neutral variation and hence may provide important signatures in sequence data of adaptation that involves clonal interference.

Our predictions for the structure of the fitness-class genealogies depend on the population size, mutation rate, and strength of selection only through the combinations $\log [ N s ]$ and $\log[U_b/s]$.  The timescales in generations are also proportional to the inverse of the strength of selection.  Thus the patterns of genetic variation in an adapting population depend only very weakly (logarithmically) on population size and mutation rate in the large-$q$ regime where clonal interference is pervasive, suggesting that there is limited power to infer these parameters from patterns of molecular evolution.  This is a consequence of the fact that the evolutionary dynamics are also only very weakly dependent on these parameters in the clonal interference regime.

We have seen that in the large-$q$ limit of our model, the ratios of the number of mutational steps to the most recent common ancestors in samples of different sizes are exactly equivalent to those expected in the Bolthausen-Sznitman coalescent \citep{bolthausensznitman98}.  This is identical to the limiting behavior of these ratios in several very different models of selection recently studied by Brunet, Derrida, and others \citep{brunetmunier06, brunetmunier07, brunetsimon08, berestycki12, brunetderrida12a}; see \citep{brunetderrida12b} for a recent review.  The reason for this equivalence between very different models remains unclear, but suggests a degree of universality: an interesting topic for future work. We emphasize, however, that the times to most recent ancestors in our model reduce to the Bolthausen-Sznitman ratios only when measured in mutational steps and only when all individuals are sampled from the same fitness class.  The ratios of time to most recent common ancestors, measured in generations, have a different form.  Nevertheless, in the limit of very large $q$, almost all the individuals will have  fitness much closer to the mean than to the nose.   As the rate of coalescence is proportional to the difference between the mean and the nose, the approximation of sampling only from the largest fitness class is asymptotically good. The modifications of the Bolthausen-Sznitman ratios  are then simply determined by adding the nose-to-mean time, (which turns out to be equal to the mean pairwise correlation time), to all the coalescent times.

Our analysis in this paper has focused on the simplest possible model of positive selection on a large number of linked sites, and we have neglected many potential complications.  For example, we have assumed that epistatic interactions between mutations can be neglected, and that the total potential supply of beneficial mutations is not significantly depleted over the course of adaptation.  This is consistent with our focus on rapidly adapting populations in the large-$q$ clonal interference regime.  As a population approaches a fitness peak, these approximations will likely fail and the dynamics of adaptation and patterns of genetic variation may either become more complex, or return to the regime where further adaptation is driven by isolated selective sweeps.  We have also focused exclusively on beneficial mutations which all have the same fitness effect $s$, and have neglected both deleterious mutations and beneficial mutations which confer different fitness effects.  This is justified by earlier work by us and others that suggests that in rapidly adapting populations, clonal interference ensures that evolution is dominated by beneficial mutations that confer a specific fitness advantage   \citep{rouzine08, nagle08, good12}.  However, we have recently analyzed the evolutionary dynamics within a population in a model which explicitly allows for a distribution of fitness effects of beneficial mutations \citep{good12}.  We and others have also analyzed the case where a mix of both beneficial and deleterious mutations are possible \citep{rouzine03, rouzine08, goyal12}.  Those works describe the variation in fitness within populations in these more complex models and hence could form the basis for a more complex version of the fitness-class coalescent method we have used here.  This generalized fitness-class coalescent would admit the possibility of mutational steps of various different sizes and towards both lower and higher fitness.

An alternative approach by one of us allows for beneficial mutations to have a variety of different effects, without making reference to fitness classes \citep{dsfjstat}.  As long as the distribution of fitness effects of potential beneficial mutations falls off faster than a simple exponential for large $s$, the dynamics in large populations is dominated by mutations with $s$ close to some value, $\tilde{s}$ \citep{good12, dsfjstat}.  In this case, most properties of the dynamics on time scales longer than the nose-to-mean time $\tsw$ are quite universal (and more strongly so when $v/\tilde{s}^2$ is large).  As $\tsw$ is also the time scale of the coalescence, this suggests that the coalescent statistics should also be universal. The continuous time results quoted above for the evolution of the frequency of a sub-population emerges naturally in this more general analysis, and indeed correspond to the universal limit of asymptotically-large populations \citep{dsfjstat}.  In the alternative regime where the distribution of fitness effects of potential beneficial mutations falls of more slowly than exponentially, mutations can jump from the bulk of the distribution to the lead.  These play an important role in the dynamics, and cause $q$ to remain small even for asymptotically large populations \citep{desaifisher07}. The behavior is then less universal, but this situation is likely to be relevant in real populations, especially in the initial stages of adaptation to a new environment.   Further study into these effects of the distribution of effects of beneficial mutations, of initial transient dynamics, and of large numbers of deleterious mutations are interesting topics for future research.

The final simplification of  our analysis is its focus on purely asexual populations: we have neglected the effects of recombination.  Thus our results are primarily applicable to interpreting the patterns of genetic variation in asexual microbial evolution experiments, though they may also be relevant to sexual organisms on short genomic distance scales within which recombination is rare on the relevant timescales.  We note however that our results provide an essential ingredient for predicting the effects of infrequent recombination on the evolutionary dynamics.  Specifically, we can use our predictions for the genetic variation between a pair of individuals sampled from the population to predict the distribution of fitnesses of recombinant offspring resulting from sex between these individuals.  This in turn determines how rare recombination alters the evolutionary dynamics and the distribution of fitnesses within the population.  It may prove possible to then in turn calculate how these shifts in evolutionary dynamics alter the patterns of genetic diversity in the population.  These extensions of our approach to analyze the effects of recombination on both evolutionary dynamics and patterns of molecular evolution are an important direction for future research.

\section{Acknowledgments}

We thank Richard Neher, Boris Shraiman, Thierry Mora, Lauren Nicolaisen, Benjamin Good, Elizabeth Jerison, and John Wakeley for many useful discussions.  MMD acknowledges support from the James S. McDonnell Foundation, the Alfred P. Sloan Foundation, and the Harvard Milton Fund. DSF acknowledges support from the National Science Foundation via DMS-1120699.

\section*{Appendix A:  Coalescence Probabilities}
In this Appendix, we carry out the calculations of coalescence probabilities in detail.  Consider  $H$ individuals who coalesce into $K$ lineages, with $h_1$ individuals coalescing into lineage $1$, $h_2$ individuals coalescing into lineage $2$, and so on, up to $h_K$ individuals coalescing into lineage $K$.  We note that $\sum_{j=1}^K h_j = H$.  We begin by asking the probability that $H$ individuals coalesce into $K$ lineages at a \emph{specific} set of $K$ sites (out of the total of $B$) in the genome: call these sites $1$ through $K$ in the genome, for concreteness.  We  also assume for now that the $H$ individuals coalesce in a \emph{specific} way into these $K$ lineages:  i.e. individual 3 coalesces into the lineage at site $5$, etc).  We denote the frequency of the lineage at site $j$ in the genome by $f_j$;  so that $f_j = \frac{\nu_j}{\sigma}$. We denote by $A$ the probability that the $H$ individuals coalesce into the $K$ lineages at these specific sites according to the specific configuration $\{ h_j \}$.

Given these definitions, we have: \eon A = \left\langle \prod_{j=1}^K f_j^{h_j} \right\rangle = \left\langle \prod_{j=1}^K \frac{\nu_j^{h_j}}{\sigma^{h_j}} \right\rangle = \left\langle \frac{1}{\sigma^H} \prod_{j=1}^K \nu_j^{h_j} \right\rangle. \eoff  We make use of the identity \eon \frac{1}{\sigma^H} = \int_0^\infty \frac{x^{H-1}}{(H-1)!} e^{-x \sigma} dx \eoff   to obtain  \eon A  =  \int_0^\infty \frac{x^{H-1}}{\Gamma(H)} \left\langle e^{-x \sigma} \prod_{j=1}^K \nu_j^{h_j} \right\rangle dx . \eoff  We now use the definition of $\sigma$ as the sum of the $\nu_j$ and separate out the $\nu_j$ that correspond to the lineages we are considering. Note that the $\nu_j$ are independent of each other.  Thus one obtains \eon A = \int_0^\infty \frac{x^{H-1}}{\Gamma(H)} \left\langle e^{-x \sum_{j=K+1}^B \nu_j} \right\rangle \left\langle \prod_{j=1}^K \nu_j^{h_j} e^{-x \nu_j} \right\rangle dx  \eoff  whence, by independence,  \eon A = \int_0^\infty \frac{x^{H-1}}{\Gamma(H)} \left\langle e^{-x \nu_1} \right\rangle^{B-K} \left\langle \prod_{j=1}^K \nu_j^{h_j} e^{-x \nu_j} \right\rangle dx .\eoff

From \eq{two} we have \eon \left\langle e^{-z \nu_i} \right\rangle = e^{-\mu_i/U z^{1-1/q}} = e^{-z^\alpha/B}, \eoff where  $\alpha \equiv 1 - \frac{1}{q}$.  Substituting this in, and assuming large $B$ so that $(B-K)/B \approx 1$, we find \eon A = \int_0^\infty \frac{x^{H-1}}{\Gamma(H)} e^{-x^\alpha} \left\langle \prod_{j=1}^K \nu_j^{h_j} e^{-x \nu_j} \right\rangle dx . \eoff  We then use that \eon \left\langle \nu^h e^{-x \nu} \right\rangle = (-1)^h \frac{\partial^h}{\partial z^h} \left[ e^{-z^\alpha/B} \right]. \eoff  Making the large-$B$ approximation that $e^{-z^\alpha/B} \approx 1 - \frac{z^\alpha}{B}$ and differentiating, we find \eon \left\langle \nu^h e^{-x \nu} \right\rangle = \frac{\alpha}{B} \frac{\Gamma(h-\alpha)}{\Gamma(1-\alpha)} x^{\alpha-h} . \eoff  Using this result, we have \eon A = \int_0^\infty \frac{x^{H-1}}{\Gamma(H)} e^{-x^\alpha} \prod_{j=1}^K \frac{\alpha x^{\alpha-h_j} \Gamma(h_j - \alpha)}{B \Gamma (1-\alpha)} dx . \eoff  Since $\sum_{j=1}^K h_j = H$ we can rewrite this as \eon A = \int_0^\infty \frac{x^{K \alpha} dx}{x} \frac{\alpha^K}{B^K \Gamma(H)} e^{-x^\alpha} \prod_{j=1}^K \frac{\Gamma(h_j - \alpha)}{\Gamma (1-\alpha)}  . \eoff  Now we define \eon y = x^\alpha \qquad dy = \alpha x^{\alpha-1} dx \qquad \frac{dy}{\alpha y} = \frac{dx}{x}, \eoff and making this change of variables obtain \eon A = \int_0^\infty \frac{dy}{\alpha} \frac{y^{K-1} e^{-y} \alpha^K}{B^K \Gamma(H)} \prod_{j=1}^K \frac{\Gamma(h_j - \alpha)}{\Gamma (1-\alpha)}  . \eoff  The $dy$ integral yields  a $\Gamma$ function, giving \eon A = \frac{\Gamma(K) \alpha^{K-1}}{B^K \Gamma(H)} \prod_{j=1}^K \frac{\Gamma(h_j - \alpha)}{\Gamma (1-\alpha)}  . \eoff

So far we have considered the probability of this coalescence event involving $K$ lineages at a specific set of $K$ sites on the genome.  We now want to sum over all the possible sets of $K$ sites on the genome at which this could occur.  There are a total of $B^K/K!$ of these.  We define $E$ to be the probability of this coalescence event involving $K$ lineages at \emph{any} set of $K$ sites on the genome.  We have \eon E = \frac{\alpha^{K-1}}{K \Gamma(H)} \prod_{j=1}^K \frac{\Gamma(h_j - \alpha)}{\Gamma (1-\alpha)}  . \eoff

Now so far we have assumed that specific individuals coalesce into specific lineages. But given a set $\{ h_j \}$ there are a total of ${H \choose h_1, h_2, \ldots h_K}$ ways to assign specific individuals to specific lineages.  Thus the total probability of $H$ individuals coalescing into $K$ lineages, in a specific configuration $\{ h_j \}$, which we will call $C_{H, K, \{h_j\}}$, is \eaon C_{H, K, \{h_j\}} & = & \frac{H!}{\prod_{j=1}^K h_j!} \frac{\alpha^{K-1}}{K \Gamma(H)} \prod_{j=1}^K \frac{\Gamma(h_j - \alpha)}{\Gamma (1-\alpha)} \\ & & \nonumber = \frac{H \alpha^{K-1}}{K} \prod_{j=1}^K \frac{\Gamma(h_j - \alpha)}{\Gamma(h_j +1) \Gamma (1-\alpha)} , \eaoff equivalent to \eq{pchk} in the main text.

To compute $\dcoal{H}{K}$, we first make the definition \eon f(H,K) = \sum_{\{ h_j \}} \prod_{j=1}^K \frac{\Gamma(h_j - \alpha)}{\Gamma(h_j +1) \Gamma (1-\alpha)} , \eoff and note that \eon \dcoal{H}{K} = \frac{H}{K \alpha} f(H,K). \eoff  There is no simple analytic expression for $f(H,K)$.  However, we can define its generating function \eon R_f(z) \equiv \sum_{H=0}^\infty f(H,K) z^H. \eoff Note we are summing from $H=0$: even though for $H<K$ this is not biologically relevant, it will be useful formally.  Now we have \eaon R_f(z) & = & \sum_{H=0}^\infty \sum_{\{h_j\}}^{constrained} f(H,K) z^H \\ & & \nonumber = \sum_{h_1=0}^\infty \sum_{h_2 = 0}^\infty \ldots \sum_{h_K=0}^\infty f(H,K) z^H . \eaoff  Substituting in for $f(H,K)$, we find \eon R_f(z) = \left[ \sum_{h=0}^\infty \frac{\alpha \Gamma(h-\alpha) z^h}{\Gamma(h+1) \Gamma(1-\alpha)} \right]^K, \eoff where we have used the fact that the sums over the different $h$ are now independent.  Recognizing the Taylor series, we have \eon R_f(z) = \left[ 1 - (1-z)^\alpha \right]^K, \eoff as quoted in the main text.  Note we can also plug in $K = 1$ to recover the result for $\ph$ quoted in \eq{ph}.

\bibliography{posselcoallib}

\end{document}